\documentclass[preprint,showpacs,nofootinbib,aps]{revtex4}
 \usepackage{graphicx}
 \usepackage{rotating}
 \usepackage{multirow}
 \usepackage{bm}
 \begin{document}
 \title{Phenomenological study of the $B_{c}$ ${\to}$ $BP$, $BV$
        decays with perturbative QCD approach}
 \author{Junfeng Sun}
 \affiliation{Institute of Particle and Nuclear Physics,
              Henan Normal University, Xinxiang 453007, China}
 \author{Yueling Yang}
 \email{yangyueling@htu.cn}
 \affiliation{Institute of Particle and Nuclear Physics,
              Henan Normal University, Xinxiang 453007, China}
 \author{Qin Chang}
 \email{changqin@htu.cn}
 \affiliation{Institute of Particle and Nuclear Physics,
              Henan Normal University, Xinxiang 453007, China}
 \author{Gongru Lu}
 \affiliation{Institute of Particle and Nuclear Physics,
              Henan Normal University, Xinxiang 453007, China}
 \begin{abstract}
 Inspired by the recent LHCb measurements and
 forthcoming great potential on $B_{c}$ meson,
 we study the exclusive $B_{c}$ ${\to}$ $B_{q}P$,
 $B_{q}V$ decays with the perturbative QCD approach,
 where $q$ $=$ $u$, $d$, $s$; $P$ and $V$ denote the lightest
 pseudoscalar and vector $SU(3)$ nonet meson, respectively.
 By retaining the quark transverse momentum,
 employing the Sudakov factors, and
 choosing the typical scale as the maximum virtualities
 of the internal particles,
 we calculate the $B_{c}$ ${\to}$ $B$ transition from factors,
 and our results show that about 90\% contribution to form
 factors come from the ${\alpha}_{s}/{\pi}$ $<$ $0.3$ region.
 The contributions of penguin and annihilation to branching
 ratios are very small due to the serious suppression by the
 CKM factors. There is some hierarchy relations among the
 $B_{c}$ ${\to}$ $BP$, $BV$ decays.
 The branching ratios for $B_{c}$ ${\to}$ $B_{d,s}{\pi}$,
 $B_{d,s}{\rho}$, $B_{s}K$ are large and could be
 measured by the running LHCb.

 \end{abstract}
 \maketitle

 \section{Introduction}
 \label{sec01}
 The $B_{c}$ meson is the heaviest ground pseudoscalar
 meson with explicit both bottom and charm flavour.
 The yield ratio of $B_{c}$ meson is very small \cite{pdg.2012},
 but it is still possible to obtain enough measurements
 to explore its property at high energy colliders.
 The $B_{c}$ meson was observed for the first time via the
 semileptonic decay $B_{c}$ ${\to}$ $J/{\psi}{\ell}{\nu}$ in
 $1.8$ TeV $p{\bar{p}}$ collisions using the CDF detector
 at the Fermilab Tevatron in 1998 \cite{prd.58.112004}.
 Recently, its mass is accurately determined at the ${\cal O}(10^{-4})$
 level from the fully reconstructed $B_{c}$ ${\to}$ $J/{\psi}{\pi}$
 mode by the CDF and LHCb experimental groups
 \cite{prl.100.182002,prl.109.232001},
 and its lifetime is also measured at the ${\sim}$ $3\%$ level by
 the LHCb collaboration \cite{arxiv:1401.6932}.

 The $B_{c}$ meson, laying below $BD$ threshold, can decay only
 via the weak interaction. Its decay modes can be divided into
 three types \cite{zpc.51.549,prd.49.3399}:
 (1) the $c$ quark decays while the $b$ quark as a spectator;
 (2) the $b$ quark decays while the $c$ quark as a spectator;
 (3) the annihilation channel.
 The $c$ quark decay modes [the type (1)] are responsible for
 about $70\%$ of the width of $B_{c}$ meson \cite{prd.53.4991}.
 This type of decay process, although very challenging to
 experiments, has recently been observed in the $B_{c}$ ${\to}$
 $B_{s}{\pi}$ mode with significance in excess of 5 standard
 deviations by the LHCb collaboration  \cite{1308.4544}.
 The $b$ quark decay modes [the type (2)] account for about
 $20\%$ of the width of $B_{c}$ meson \cite{0412158}. The
 $b$ ${\to}$ $c$ transition offers a well-reconstructed
 experimental signature at the Tevatron and LHC,
 for example, in the decay modes of
 $B_{c}^{+}$ ${\to}$ $J/{\psi}{\pi}^{+}$
 \cite{prl.100.182002,prl.109.232001,prl.101.012001},
 ${\psi}(2S){\pi}^{+}$ \cite{1303.1737},
 $J/{\psi}D_{s}^{(\ast)+}$ \cite{prd.87.112012},
 $J/{\psi}K^{+}K^{-}{\pi}^{+}$ \cite{1309.0587},
 $J/{\psi}{\pi}^{+}{\pi}^{-}{\pi}^{+}$ \cite{1204.0079},
 $J/{\psi}e^{+}{\nu}_{e}$ \cite{prl.97.012002}
 and so on.
 The weak annihilation mode [the type (3)] is estimated
 to take $10\%$ shares of the width of $B_{c}$ meson
 \cite{0412158}. The pure weak annihilation decay to
 two light mesons, $B_{c}$ ${\to}$ $u$ $+$ $d$, is so highly
 helicity-suppressed that there is little probability of
 detecting the charmless and/or bottomless hadronic
 decays $B_{c}$ ${\to}$ $PP$, $PV$, $VV$ \cite{prd.81.074012},
 where $P$ and $V$ denote the lightest $SU(3)$ pseudoscalar
 and vector mesons, respectively;
 and to date, no corresponding measurements exist.

 It is estimated that one could expect ${\cal O}(10^{10})$
 of $B_{c}$ mesons per year at the LHC \cite{pan.67.1559}.
 Along with the running of the LHC, more and more $B_{c}$
 decay modes will be observed.
 Anticipating the experimental developments, many studies
 (see Table.\ref{tab01}) have been devoted to the bottom
 conserving and charm changing decay modes $B_{c}$ ${\to}$
 $BP$, $BV$, including estimates undertaken within various
 quark models assisted by confining potential \cite{prd.86.094028,
 prd.80.114003,prd.74.074008,prd.73.054024,epjc.32.29},
 with potential models based on the Bethe-Salpeter equation
 \cite{prd.49.3399,prd.62.014019},
 with BSW or ISGW models \cite{prd.39.1342,zpc.51.549},
 with QCD sum rules \cite{pan.67.1559},
 with heavy quark spin symmetry \cite{prd.61.034012},
 with QCD factorization at the leading order \cite{prd.77.114004},
 but without perturbative QCD (pQCD) approach.
 In this paper, we study the $B_{c}$ ${\to}$ $BP$, $BV$
 decays with the pQCD approach \cite{pqcd} to fill in this
 gap and provide a ready reference to the existing and
 forthcoming experiments.

 This paper is organized as follows:
 In Section \ref{sec2}, we discuss the theoretical framework,
 compute the $B_{c}$ ${\to}$ $B$ transition form factors and
 the amplitudes for $B_{c}$ ${\to}$ $BP$, $BV$ decays with
 the pQCD approach.
 The section \ref{sec3} is devoted to the numerical results.
 Finally, we summarize in Section \ref{sec4}.
 \section{Theoretical framework and the decay amplitudes}
 \label{sec2}
 \subsection{the effective Hamiltonian}
 \label{sec21}
 Because of the hierarchy $m_{W^{\pm}}$ ${\gg}$ $m_{b,c}$
 ${\gg}$ ${\Lambda}_{\rm QCD}$ (where $m_{W^{\pm}}$ and
 $m_{b,c}$ are the mass of the $W^{\pm}$ boson and $b$,
 $c$ quarks, respectively; ${\Lambda}_{\rm QCD}$ is the
 QCD confinement scale),
 one typically use the effective field theory to deal with
 weak decays of the hadron containing heavy quark.
 Using the operator product expansion, the low energy
 effective Hamiltonian relevant to nonleptonic $B_{c}$
 ${\to}$ $BP$, $BV$ decays can be written as \cite{9512380}:
  \begin{eqnarray}
 {\cal H}_{\rm eff} &=&
  \frac{G_{F}}{\sqrt{2}} \Big\{
  V_{ub}V_{cb}^{\ast}
  \Big[C_{1}^{a}({\mu})Q_{1}^{a}({\mu})+C_{2}^{a}({\mu})Q_{2}^{a}({\mu})\Big]
  \nonumber \\ &+&
  \sum\limits_{q_{_{1}}\!,q_{_{2}}}V_{uq_{_{1}}}\!V_{cq_{_{2}}}^{\ast}
  \Big[C_{1}({\mu})Q_{1}({\mu})+C_{2}({\mu})Q_{2}({\mu})\Big]
  \nonumber \\ &+&
  \sum\limits_{q_{_{3}}}\sum\limits_{k=3}^{10}
  V_{uq_{_{3}}}\!V_{cq_{_{3}}}^{\ast}C_{k}({\mu})Q_{k}({\mu})
  \Big\}+\text{h.c.}
  \label{Hamiltonian},
  \end{eqnarray}
 where $G_{F}$ is the Fermi coupling constant;
 $q_{i}$ denotes the down-type quarks $d$ and $s$.
 The Wilson coefficients $C_{i}({\mu})$ summarize the
 contributions from scales higher than ${\mu}$, which
 are calculable and can be evaluated to the scale
 ${\mu}$ with the renormalization group equation.
 Their numerical values at four different scales ${\mu}$ are
 listed in Table.\ref{tab02}.
 The expressions of the local four-quark operators $Q_{i}$
 can be written explicitly as follows:
 \begin{itemize}
 \item[(i)] current-current (tree) operators
   \begin{eqnarray}
   Q^{a}_{1}&=&(\bar{b}_{\alpha}c_{\alpha})_{V-A}
               (\bar{u}_{\beta}b_{\beta})_{V-A}
   \label{q1a}, \\
   Q^{a}_{2}&=&(\bar{b}_{\alpha}c_{\beta})_{V-A}
               (\bar{u}_{\beta}b_{\alpha})_{V-A}
   \label{q2a}, \\
   Q_{1}&=&(\bar{q}_{2\alpha}c_{\alpha})_{V-A}
           (\bar{u}_{\beta}q_{1\beta})_{V-A}
   \label{q1}, \\
   Q_{2}&=&(\bar{q}_{2\alpha}c_{\beta})_{V-A}
           (\bar{u}_{\beta}q_{1\alpha})_{V-A}
   \label{q2},
   \end{eqnarray}
 \item[(ii)] QCD penguin operators
   \begin{eqnarray}
   Q_{3}&=& \sum\limits_{q}
           (\bar{u}_{\alpha}c_{\alpha})_{V-A}
           (\bar{q}_{\beta}q_{\beta})_{V-A}
   \label{q3}, \\
   Q_{4}&=& \sum\limits_{q}
           (\bar{u}_{\alpha}c_{\beta})_{V-A}
           (\bar{q}_{\beta}q_{\alpha})_{V-A}
   \label{q4}, \\
   Q_{5}&=& \sum\limits_{q}
           (\bar{u}_{\alpha}c_{\alpha})_{V-A}
           (\bar{q}_{\beta}q_{\beta})_{V+A}
   \label{q5}, \\
   Q_{6}&=& \sum\limits_{q}
           (\bar{u}_{\alpha}c_{\beta})_{V-A}
           (\bar{q}_{\beta}q_{\alpha})_{V+A}
   \label{q6},
   \end{eqnarray}
 \item[(iii)] electroweak penguin operators
   \begin{eqnarray}
   Q_{7}&=& \sum\limits_{q} \frac{3}{2}Q_{q}
           (\bar{u}_{\alpha}c_{\alpha})_{V-A}
           (\bar{q}_{\beta}q_{\beta})_{V+A}
   \label{q7}, \\
   Q_{8}&=& \sum\limits_{q} \frac{3}{2}Q_{q}
           (\bar{u}_{\alpha}c_{\beta})_{V-A}
           (\bar{q}_{\beta}q_{\alpha})_{V+A}
   \label{q8}, \\
   Q_{9}&=& \sum\limits_{q} \frac{3}{2}Q_{q}
           (\bar{u}_{\alpha}c_{\alpha})_{V-A}
           (\bar{q}_{\beta}q_{\beta})_{V-A}
   \label{q9}, \\
   Q_{10}&=& \sum\limits_{q} \frac{3}{2}Q_{q}
           (\bar{u}_{\alpha}c_{\beta})_{V-A}
           (\bar{q}_{\beta}q_{\alpha})_{V-A}
   \label{q10},
   \end{eqnarray}
 \end{itemize}
 where the tree operators of $Q^{a}_{1,2}$ describe the weak
 annihilation topology; ${\alpha}$ and ${\beta}$ are the color
 indices; The $q$ in penguin operators denotes all the active
 quarks at scale ${\mu}$ $=$ ${\cal O}(m_{c})$, i.e. $q$ $=$
 $u$, $d$, $s$, $c$; The left- and right-handed currents are
 defined as $(\bar{q}_{\alpha}q_{\beta}^{\prime})_{V{\pm}A}$
 ${\equiv}$ $\bar{q}_{\alpha}{\gamma}_{\mu}(1{\pm}{\gamma}_{5})
 q_{\beta}^{\prime}$; and $Q_{q}$ is the charge of quark $q$
 in the unit of ${\vert}e{\vert}$.
 \subsection{Hadronic matrix elements}
 \label{sec22}
 The essential problem obstructing the calculation of decay
 amplitude is how to properly evaluate the hadronic matrix
 elements of the local operators.
 Using the Brodsky-Lepage approach \cite{prd.22.2157},
 the hadronic matrix elements can be written as the
 convolution of a hard-scattering kernels containing
 perturbative QCD contributions with the universal
 wave functions reflecting the nonperturbative dynamics.
 Currently, there are three popular phenomenological
 approaches to evaluate the hadronic matrix elements
 as an expansion in the strong coupling constant
 ${\alpha}_{s}$ and in the ratio ${\Lambda}_{\rm QCD}/m_{Q}$,
 which are entitled to
 QCD factorization (QCDF) \cite{qcdf},
 the soft-collinear effective theory (SCET) \cite{scet},
 and the pQCD approach \cite{pqcd}.
 These methods differ from each other in several
 aspects. For example, only the collinear degrees of
 freedom are taken into account in QCDF and SCET,
 while the transverse momenta implemented with the
 help of the Sudakov formalism in pQCD approach.
 The other different features of these methods are
 power counting, the choice of the scale at which
 the strong interaction effects are calculated,
 how to deal with the contribution of spectator
 scattering and weak annihilation, and so on.
 With the running LHCb and the advent of SuperKEKB
 physics program, the precision of observables will
 be greatly improved, and it should be possible to
 disentangle the underlying dynamics in nonleptonic
 $B$ decays.

 In this paper, we study the $B_{c}$ ${\to}$ $BP$, $BV$
 decays with the pQCD approach.
 By keeping the parton transverse momentum and
 employing the Sudakov factors to modify the endpoint
 behavior, the hadron matrix elements are expressed as
 the convolution of wave functions and the heavy quark
 decay subamplitudes, integrated over the longitudinal
 and transverse momenta.
 After the Fourier transformation, the typical formula
 of the hadron matrix elements can be written as:
 \begin{eqnarray}
 M&{\propto}&
 {\int}{\bf d}x_{1}{\bf d}x_{2}{\bf d}x_{3}
 {\int}{\bf d}\vec{b}_{1}{\bf d}\vec{b}_{2}
 {\bf d}\vec{b}_{3}
 {\phi}_{B_{c}}(x_{1},\vec{b}_{1})
 {\phi}_{B_{q}}(x_{2},\vec{b}_{2})
  \nonumber \\ & & {\times}
 {\phi}_{P,V}(x_{3},\vec{b}_{3})
 e^{-S_{B_{c}}(t)-S_{B_{q}}(t)-S_{P,V}(t)}
 H(x_{i},\vec{b}_{i},t)
 \label{hme01},
 \end{eqnarray}
 where ${\phi}_{i}$ is the meson wave functions;
 $\vec{b}_{i}$ is the conjugate variable of the transverse
 moment $\vec{k}_{i{\perp}}$ of valence quark;
 $e^{-S_{i}(t)}$ is the Sudakov factor;
 $H$ is the process-dependent heavy quark decay subamplitudes.
 The kinematic variables and wave functions are given
 as below.
 \subsection{kinematic variables}
 \label{sec23}
 In the terms of the light cone coordinate,
 the momenta of the valence quarks and hadrons in the
 rest frame of the $B_{c}$ meson are defined as:
  \begin{eqnarray}
  p_{1}&=&\frac{m_{1}}{\sqrt{2}}(1,1,0),
  \label{p1}, \\
  p_{2}&=&(q_{2}^{+},q_{2}^{-},0)
  \label{p2}, \\
  p_{3}&=&(q_{3}^{-},q_{3}^{+},0)
  \label{p3}, \\
  k_{i}&=&x_{i}p_{i}+(0,0,\vec{k}_{i{\perp}})
  \label{ki}, \\
 {\epsilon}_{\parallel}&=&\frac{1}{m_{3}}(-q_{3}^{-},q_{3}^{+},0)
  \label{e3}, \\
  q_{i}^{\pm}&=&\frac{E_{i}{\pm}p}{\sqrt{2}}
  \label{si},
  \end{eqnarray}
  where the subscript $i$ $=$ $1$, $2$, $3$ refers to
  $B_{c}$, $B_{q}$ and the light meson, respectively;
  $k_{i}$, $\vec{k}_{i{\perp}}$, $x_{i}$ are the momentum,
  transverse momentum and longitudinal momentum fraction
  of light valence quark confined within meson,
  respectively;
  ${\epsilon}_{\parallel}$ denotes the longitudinal
  polarization vector of the light vector meson.
  $E_{i}$ and $p$ are the energy and the momentum
  of final state, respectively.
  For the sake of brevity, the Lorentz-invariant variables
  are defined by
  \begin{equation}
  s=2p_{2}{\cdot}p_{3},~~
  t=2p_{1}{\cdot}p_{2},~~
  u=2p_{1}{\cdot}p_{3}
  \label{stu}.
  \end{equation}
  \subsection{wave functions}
  \label{sec24}
  In order to get the analytic formulas of the decay
  amplitudes, we use the light-cone wave functions
  which can be decomposed as \cite{prd.65.014007}:
  \begin{eqnarray}
 {\langle}0{\vert}\bar{b}_{\alpha}(0)c_{\beta}(z)
 {\vert}B_{c}(p_{1}){\rangle}
 &=&
  \frac{-if_{B_{c}}}{4N_{c}}{\int}{\bf d}^{4}k_{1}
  \Big\{ e^{-ik_{1}{\cdot}z}{\phi}_{B_{c}}
  \big(\!\!\not{\!p}_{1}\!+\!m_{B_{c}}\!\big)
 {\gamma}_{5}\Big\}_{{\beta}{\alpha}}
  \label{wfbc-01}, \\
 {\langle}B_{q}(p_{2}){\vert}
  \bar{q}_{\alpha}(z)b_{\beta}(0){\vert}0{\rangle}
 &=&
  \frac{-if_{B_{q}}}{4N_{c}}{\int}{\bf d}^{4}k_{2}
  \Big\{ e^{+ik_{2}{\cdot}z}{\phi}_{B_{q}}
 {\gamma}_{5}
  \big(\!\!\not{\!p}_{2}\!+\!m_{B_{q}}\!\big)
  \Big\}_{{\beta}{\alpha}}
  \label{wfbq-01}, \\
 {\langle}P(p_{3}){\vert}
  \bar{q}_{1\alpha}(z)q_{2\beta}(0){\vert}0{\rangle}
 &=&
  \frac{-if_{P}}{4N_{c}}
 {\int}{\bf d}^{4}k_{3}\,e^{+ik_{3}{\cdot}z}
  \Big\{ {\gamma}_{5}\big[
  \!\!\not{\!p}_{3}{\phi}_{P}^{a}\!
 +\!{\mu}_{P}{\phi}_{P}^{p}\!
 +\!{\mu}_{P}(\!\not{\!n}_{-}\!\!\not{\!n}_{+}\!-\!1){\phi}_{P}^{t}
  \big] \! \Big\}_{{\beta}{\alpha}}
  \label{wfp-01}, \\
 {\langle}V(p_{3},{\epsilon}_{\parallel}){\vert}
  \bar{q}_{1\alpha}(z)q_{2\beta}(0){\vert}0{\rangle}
 &=&
  \frac{f_{V}}{4N_{c}}
 {\int}{\bf d}^{4}k_{3}\,{\rm e}^{+ik_{2}{\cdot}z}
  \Big\{ \!\!\not{\!\epsilon}_{\parallel}
  \big[m_{V}{\phi}_{V}
  +\!\not{\!p}_{3}\frac{f_{V}^{T}}{f_{V}}{\phi}_{V}^{t}\big]
  +\frac{m_{V}f_{V}^{T}}{f_{V}}{\phi}_{V}^{s}\Big\}_{{\beta}{\alpha}}
  \label{wfv-01},
  \end{eqnarray}
 where $N_{c}$ $=$ 3 is the color number;
 $f_{i}$ is the decay constant.
 The explicit expressions of the light-cone distribution
 amplitudes (${\phi}_{B_{c}}$, ${\phi}_{B_{q}}$,
 ${\phi}_{P}^{a,p,t}$,${\phi}_{V}$ and ${\phi}_{V}^{t,s}$)
 are collected in Appendix. \ref{app01} and \ref{app02}.
  \subsection{Form factor}
  \label{sec25}
  The $B_{c}$ ${\to}$ $B_{q}$ form factors are defined
  as \cite{bsw}:
  \begin{equation}
 {\langle}B_{q}(p_{2}){\vert}(\bar{q}c)^{\mu}_{V-A}
 {\vert}B_{c}(p_{1}){\rangle}
 =\Big\{(p_{1}+p_{2})^{\mu}
 -\frac{m_{1}^{2}-m_{2}^{2}}{q^{2}}q^{\mu}\Big\}F_{1}
 +\frac{m_{1}^{2}-m_{2}^{2}}{q^{2}}q^{\mu}F_{0}
  \label{formfactor01}
  \end{equation}
 where $q$ $=$ $p_{1}$ $-$ $p_{2}$ is the momentum
 transfer.
 Usually, the longitudinal form factor $F_{0}(q^{2})$
 is compulsorily equal to the transverse form factor
 $F_{1}(q^{2})$ in the largest recoil limit to cancel
 singularities appearing at the pole $q^{2}$ $=$ $0$,
 i.e., $F_{0}(0)$ $=$ $F_{1}(0)$.

 The $B_{c}$ ${\to}$ $B$ transition form factors
 can be written as the convolution of wave functions
 and the one-gluon exchange scattering amplitudes
 using the pQCD approach.
 There are two types of diagrams contributing to the
 $B_{c}$ ${\to}$ $B$ transition form factors,
 which are displayed in Fig.\ref{fig01}.
 The expression of the form factors are written as
 \begin{eqnarray}
  F_{1}(q^{2})&=&
  \frac{{\pi}C_{F}}{N_{c}}f_{B_{c}}f_{B_{q}}
 {\int}_{0}^{1}{\bf d}x_{1}{\bf d}x_{2}
 {\int}_{0}^{\infty}{\bf d}b_{1}{\bf d}b_{2}
 {\phi}_{B_{c}}(x_{1})
 {\phi}_{B_{q}}(x_{2},b_{2})
  \nonumber \\ &{\times}&
  \Big\{H_{a}
  \big[\{m_{1}(2m_{2}-m_{1})+q^{2}\}x_{2}
  +m_{c}(2m_{1}-m_{2})-q^{2}\big]
  \nonumber \\ & &+
  H_{b}
  \big[\{m_{2}(2m_{1}-m_{2})+q^{2}\}x_{1}-q^{2}\big]
  \Big\}
  \label{formfactor02},
  \end{eqnarray}
  \begin{eqnarray}
  F_{0}(q^{2})&=&
  \frac{{\pi}C_{F}}{N_{c}}f_{B_{c}}f_{B_{q}}
 {\int}_{0}^{1}{\bf d}x_{1}{\bf d}x_{2}
 {\int}_{0}^{\infty}{\bf d}b_{1}{\bf d}b_{2}
 {\phi}_{B_{c}}(x_{1})
 {\phi}_{B_{q}}(x_{2},b_{2})
  \nonumber \\ &{\times}&
  \Big\{H_{a}
  \big[\{(m_{1}-m_{2})^{2}+m_{2}^{2}-q^{2}\}x_{2}
  \nonumber \\ & &~~~~~~+
   2m_{2}(2m_{1}-m_{2})
  - m_{c}(2m_{1}+m_{2})
  + q^{2}\big]
  \nonumber \\ & &-
  H_{b}
  \big[\{(m_{1}-m_{2})^{2}+m_{1}^{2}-q^{2}\}x_{1}
  \nonumber \\ & &~~~~~~+
   2m_{1}(2m_{2}-m_{1})+q^{2}\big]
  \Big\} \frac{q^{2}}{m_{1}^{2}-m_{2}^{2}}
  +F_{1}(q^{2})
  \label{formfactor03}.
  \end{eqnarray}

  It is well known that the $q^{2}$-dependent behavior
  of the form factor is required in semileptonic $B_{c}$
  decays. To shed light on the momentum dependence, one
  needs a specific model to parameterize the form factors.
  Here we adopt the three-parameter form, i.e.
  \begin{equation}
  F_{i}(q^{2})=\frac{F_{i}(0)}{1-
   \displaystyle \frac{q^{2}}{m^{2}}+
  {\delta}\frac{q^{4}}{m^{4}} }
  \label{formfactor04},
  \end{equation}
  where the pole mass $m$ and curvature parameter ${\delta}$
  can be given by fit data of $q^{2}$-dependent form factors.

  \subsection{decay amplitudes and branching ratios}
  \label{sec26}
  There are generally eight diagrams (see Fig.\ref{fig02})
  contributing to the $B_{c}$ ${\to}$ $BP$, $BV$
  decays at the lowest order with the pQCD approach.
  For example, the amplitude of the $B_{c}$ ${\to}$ $B_{s}K$
  decay can be written as:
  \begin{eqnarray}
  {\cal A}(B_{c}^{+}{\to}B_{s}^{0}K^{+})
   &=&V_{us}V_{cs}^{\ast}\{
      a_{1}M^{P}_{ab,1}
     +C_{2}M^{P}_{cd,1} \}
    - V_{ub}V_{cb}^{\ast}\{
      (a_{4}-a_{10}/2)M^{P}_{ab,1}
   \nonumber \\ &+&
      (a_{6}-a_{8}/2)M^{P}_{ab,3}
     +(C_{3}-C_{9}/2)M^{P}_{cd,1}
     +(C_{5}-C_{7}/2)M^{P}_{cd,3}
   \nonumber \\ &-&
      a_{1}M^{P}_{ef,1}
     -C_{2}M^{P}_{gh,1} \}
   \label{example},
   \end{eqnarray}
  where $V_{us}V_{cs}^{\ast}$ and $V_{ub}V_{cb}^{\ast}$ are
  the CKM factors;
  $C_{i}$ are the Wilson coefficients; the parameters
  $a_{i}$ are defined as:
  \begin{eqnarray}
   a_{i}&=&C_{i}+C_{i+1}/N_{c},\quad
   (i=1,3,5,7,9)
   \label{aiodd}, \\
   a_{i}&=&C_{i}+C_{i-1}/N_{c},\quad
   (i=2,4,6,8,10)
   \label{aieven}.
   \end{eqnarray}
  The $M_{ab}$, $M_{cd}$, $M_{ef}$, $M_{gh}$ denote the
  contributions of the factorizable emission diagrams
  [Fig.\ref{fig02} (a,b)], the nonfactorizable emission
  diagrams [Fig.\ref{fig02} (c,d)], the factorizable
  annihilation diagrams [Fig.\ref{fig02} (e,f)],
  the nonfactorizable annihilation diagrams
  [Fig.\ref{fig02} (g,h)], respectively.
  They are defined as
  \begin{eqnarray}
   M^{P,V}_{ab,i}&=&M^{P,V}_{a,i}+M^{P,V}_{b,i}, \quad
   M^{P,V}_{cd,i} =(M^{P,V}_{c,i}+M^{P,V}_{d,i})/N_{c}
   \label{ampcd}, \\
   M^{P,V}_{ef,i}&=&M^{P,V}_{e,i}+M^{P,V}_{f,i},\quad
   M^{P,V}_{gh,i} =(M^{P,V}_{g,i}+M^{P,V}_{h,i})/N_{c}
   \label{ampgh}.
   \end{eqnarray}
 Here the superscripts $P$ and $V$ on $M^{P,V}$ mean that
 the light final states are the pseudoscalar and vector
 mesons, respectively; the subscript $i$ on $M_{i,j}$
 corresponds to one index of Fig.\ref{fig02};
 the subscript $j$ on $M_{i,j}$ refers to one of
 three possible Dirac structures, namely $j$ $=$ $1$ for
 $(V-A){\otimes}(V-A)$,
 $j$ $=$ $2$ for $(V-A){\otimes}(V+A)$,
 $j$ $=$ $3$ for $-2(S-P){\otimes}(S+P)$.
 The expressions of these building blocks $M_{i,j}^{k}$
 of amplitudes are displayed in Appendix.\ref{app03}.
 Our study show that
 (1) for the factorizable topologies
 [Fig.\ref{fig02} (a,b,e,f)], the contribution of the
 color-singlet-current operators
  $(\bar{q}_{1{\alpha}}q_{2{\alpha}})_{j}
   (\bar{q}_{3{\beta}}q_{4{\beta}})_{j}$
 is $N_{c}$ times larger than that of the corresponding
 color-current operators
  $(\bar{q}_{1{\alpha}}q_{2{\beta}})_{j}
   (\bar{q}_{3{\beta}}q_{4{\alpha}})_{j}$;
 (2) for the nonfactorizable topologies [Fig.\ref{fig02}
 (c,d,g,h)], the color-singlet-current operators
 contribute nothing.
 (3) The nonfactorizable contributions corresponding
 to terms of both $M_{cd,i}^{P,V}$ and $M_{gh,i}^{P,V}$
 are color-suppressed relative to the factorizable
 contributions corresponding to terms of both
 $M_{ab,i}^{P,V}$ and $M_{ef,i}^{P,V}$.
 (4) The nonfactorizable contributions might be important
 for the $B_{c}$ ${\to}$ $B_{u}P$, $B_{u}V$ decays, where
 term $M_{cd,1}^{P,V}$ is always multiplied by the large
 Wilson coefficient $C_{1}$.

 As for the mixing of physical states ${\eta}$ and
 ${\eta}^{\prime}$ meson, they are usually expressed as
 a linear combination of states in either $SU(3)$
 octet-singlet or quark-flavor mixing scheme.
 We will adopt the quark-flavor basis description
 proposed in \cite{prd.58.114006}, i.e.
   \begin{equation}
   \left(\begin{array}{c}
  {\eta} \\ {\eta}^{\prime}
   \end{array}\right) =
   \left(\begin{array}{cc}
  {\cos}{\phi} & -{\sin}{\phi} \\
  {\sin}{\phi} &  {\cos}{\phi}
   \end{array}\right)
   \left(\begin{array}{c}
  {\eta}_{q} \\ {\eta}_{s}
   \end{array}\right)
   \label{mixing01},
   \end{equation}
  where ${\eta}_{q}$ $=$ $(u\bar{u}+d\bar{d})/{\sqrt{2}}$
  and ${\eta}_{s}$ $=$ $s\bar{s}$, respectively;
  the mixing angle ${\phi}$ $=$ $(39.3{\pm}1.0)^{\circ}$
  \cite{prd.58.114006}.
  We assume that the distribution amplitudes of ${\eta}_{q}$
  and ${\eta}_{s}$ are the same as those of ${\pi}$ meson,
  but with different decay constants and chiral parameters
   \cite{prd.58.114006,prd.76.074018},
   \begin{eqnarray}
  f_{q}&=&(1.07{\pm}0.02)f_{\pi}
   \label{fetaq}, \\
  f_{s}&=&(1.34{\pm}0.06)f_{\pi}
   \label{fetas}, \\
  {\mu}_{{\eta}_{q}}&=&
   \frac{m_{{\eta}_{q}}^{2}}{m_{u}+m_{d}}
   \label{chiral-etaq}, \\
  {\mu}_{{\eta}_{s}}&=&
   \frac{m_{{\eta}_{s}}^{2}}{2m_{s}}
   \label{chiral-etas},
   \end{eqnarray}
   \begin{eqnarray}
   m_{{\eta}_{q}}^{2}&=& \displaystyle
   m_{\eta}^{2}{\cos}^{2}{\phi}
  +m_{{\eta}^{\prime}}^{2}{\sin}^{2}{\phi}
  -\frac{\sqrt{2}f_{s}}{f_{q}}
  (m_{{\eta}^{\prime}}^{2}- m_{\eta}^{2})
  {\cos}{\phi}\ {\sin}{\phi}
   \label{metaq}, \\
   m_{{\eta}_{s}}^{2}&=& \displaystyle
   m_{\eta}^{2}{\sin}^{2}{\phi}
  +m_{{\eta}^{\prime}}^{2}{\cos}^{2}{\phi}
  -\frac{f_{q}}{\sqrt{2}f_{s}}
  (m_{{\eta}^{\prime}}^{2}- m_{\eta}^{2})
  {\cos}{\phi}\ {\sin}{\phi}
   \label{metas}.
   \end{eqnarray}

  The gluonic contributions are not considered in our
  calculation, because it is shown that (1) the fraction
  of gluonium contributions to ${\eta}$ and ${\eta}^{\prime}$
  is less than 15\% \cite{JHEP.05.006.2007};
  (2) the flavor-singlet contributions from the gluonic content
  of ${\eta}^{({\prime})}$ meson is very small and can be
  neglected safely \cite{prd.87.097501}.
  In addition, the contributions from the possible $c\bar{c}$
  compositions of ${\eta}^{({\prime})}$ meson is also not
  considered here.

  In contrast, we assume the vector mesons are ideally mixed,
  i.e. the ${\omega}$ $=$ $(u\bar{u}+d\bar{d})/\sqrt{2}$
  and ${\phi}$ $=$ $s\bar{s}$.
  In fact, the $B_{c}$ ${\to}$ $B{\phi}$ decay is
  forbidden by the kinematic constrain because the $B_{c}$
  meson is below the $B{\phi}$ threshold.
  So there is a total of seventeen $B_{c}$ ${\to}$ $BP$,
  $BV$ decay modes. The decay amplitudes are listed
  in Appendix. \ref{app04}.
  The branching ratio in the $B_{c}$ meson rest frame can
  be written as
   \begin{equation}
  {\cal B}r(B_{c}{\to}BM)=
   \frac{G_{F}^{2}{\tau}_{B_{c}}}{16{\pi}}
   \frac{p}{m_{B_{c}}^{2}}
  {\vert}{\cal A}(B_{c}{\to}BM){\vert}^{2}
   \label{br01},
   \end{equation}
  where the lifetime of $B_{c}$ meson is
  ${\tau}_{B_{c}}$ $=$ $0.453{\pm}0.041$ ps
  \cite{pdg.2012}.
  \section{Numerical results and discussions}
  \label{sec3}
  The form factor and branching ratio depend on many parameters.
  To be specific, the parameters used in our calculation are
  listed in Table.\ref{tab03}. If not specified explicitly,
  we will take their central values as the default input.
  At the beginning of calculation, we would like to claim that
  we have no intention to claim a precise prediction, but to
  provide an order of magnitude estimation in order to test
  the applicability of the pQCD approach for the $B_{c}$
  ${\to}$ $BP$, $BV$ decays.

  Our numerical results on the form factors are given in Table.
  \ref{tab04}, where the uncertainties come from the mass
  $m_{b}$ $=$ $4.18{\pm}0.03$ GeV for $b$ quark,
  $m_{c}$ $=$ $1.275{\pm}0.025$ GeV for $c$ quark,
  shape parameters of distribution amplitudes, i.e.
  ${\omega}_{B_{c}}$ $=$ $0.50{\pm}0.05$ GeV for $B_{c}$ meson,
  ${\omega}_{B_{q}}$ $=$ $0.45{\pm}0.05$ ($0.55{\pm}0.05$) GeV
  for $B_{u,d}$ ($B_{s}$) meson, and the typical scale
  $(1{\pm}0.1)t$, respectively.

  There are some comments on the form factors.

  (1)
  The isospin is a good symmetry for the form
  factor $F_{0,1}^{B_{c}{\to}B_{u}}$ $=$
  $F_{0,1}^{B_{c}{\to}B_{d}}$, including the fitted pole
  mass $m$ and curvature parameter ${\delta}$.
  Considering the uncertainties, the values of form
  factors $F_{0,1}^{B_{c}{\to}B_{q}}$ at the pole
  $q^{2}$ $=$ $0$ are consistent with the recent results
  estimated with the relativistic independent quark model,
  where $F_{0,1}^{B_{c}{\to}B_{u,d}}(0)$ $=$ $1.01$
  and $F_{0,1}^{B_{c}{\to}B_{s}}(0)$ $=$ $1.03$
  \cite{prd.86.094028}. As it is well known,
  the spectator is the heavy $b$ quark in the
  $B_{c}$ ${\to}$ $B$ transition.
  The velocity of the $B$ meson is very low
  in the rest frame of the $B_{c}$ meson.
  The wave functions of $B_{c}$ and $B$ mesons
  overlap severely, which result in the
  large $B_{c}$ ${\to}$ $B$ transition form factors.

  (2)
  The $q^{2}$ dependence of the form factor is
  displayed in Fig.\ref{fig03}.
  From Eq.(\ref{formfactor03}), we can see that the
  interference between Fig.\ref{fig01}(a) and (b) is
  destructive to $F_{0}(q^{2})$ $-$ $F_{1}(q^{2})$,
  so the shape line of $F_{0}(q^{2})$ via $q^{2}$
  should be close to that of $F_{1}(q^{2})$.
  The shape lines will go up slowly at the beginning part,
  due to that with the increasing $q^{2}$,
  the velocity of the $B$ meson become much low
  which leads to serious overlap between the wave
  functions of $B_{c}$ and $B_{q}$ mesons.
  But the shape lines will go down for large $q^{2}$,
  because the form factor $F_{1}(q^{2})$ reduces
  with increasing $q^{2}$ [see Eq.(\ref{formfactor02})].

  (3) The form factors are sensitive to the choice of
  the shape parameter ${\omega}_{B_{q}}$ and the scale.
  In addition, the uncertainties from the decay constants
  of $f_{B_{c}}$ and $f_{B_{q}}$ are small,
  about 1\% and 2\%, respectively.

  (4) The contributions to form factor
  $F_{0}^{B_{c}{\to}B_{s}}(0)$ from different region of
  ${\alpha}_{s}/{\pi}$ is displayed in Fig.\ref{fig04},
  where $e^{-S}$ ${\neq}$ $1$ ($=$ $1$) denote results
  with (without) the Sudakov factor;
  $b_{i}$ is the conjugate variable of the
  transverse moment $k_{i{\perp}}$;
  ${\alpha}$ [see Eq.(\ref{alphae})] and
  ${\beta}$ [see Eq.(\ref{betaa}) and Eq.(\ref{betab})]
  are the virtuality of the internal gluon and quark,
  respectively.
  From Fig.\ref{fig04}(a) we can see that if one choose
  the virtuality of the internal gluon and quark as the
  typical scale, the contribution to form factor
  from ${\alpha}_{s}/{\pi}$ $<$ $0.3$ region is less than 40\%,
  that is to say, the hard and soft contributions to the
  form factor have the same behavior. This is the QCDF's
  viewpoint of that the form factor is not fully calculable
  in the hard scattering picture with the perturbation
  theory and that the form factor should be regarded as
  a nonperturbative quantity \cite{qcdf}.
  From Fig.\ref{fig04}(b) we can see that by keeping the
  quark transverse momentum $k_{T}$, and employing the
  Sudakov factors to suppress the kinematic configuration
  when both longitudinal and transverse momentum are soft,
  the contribution to form factor from
  ${\alpha}_{s}/{\pi}$ $<$ $0.3$ region is about 90\% and
  the percentage of contribution from large
  ${\alpha}_{s}/{\pi}$ region is small.
  Our study also shows that besides retaining the quark
  transverse momentum $k_{\perp}$ to smear the endpoint
  divergence behavior and using the Sudakov factor to
  suppress the nonperturbative contribution in large $b$
  region \cite{pqcd}, as the discussion in \cite{epjc.28.515},
  the choice of the hard scale is one of the important
  ingredients of the pQCD approach, which deserve much attention.
  If the scale $t$ is chosen as Eq.(\ref{tab}),
  then it shows that most of the contributions come
  from the ${\alpha}_{s}/{\pi}$ $<$ 0.3 region,
  implying that the pQCD approach is applicable to
  the $B_{c}$ ${\to}$ $B$ transition form factors.
  Of course, there are some controversies, even suspicion,
  about the suppression mechanism of the Sudakov factor
  on the nonperturbative contribution, about the choice
  of the hard scale and so on. The deeper discussion of
  these problems is needed and should be preformed, but
  beyond the scope of this paper.

  Our numerical results on the branching ratios are given
  in Table.\ref{tab05}, where the explanation of uncertainties
  is the same as that for form factors in Table.\ref{tab04}.
  There are some comments on the branching ratios.

  (1)
  From Table.\ref{tab01}, we can see that different branching
  ratios  of $B_{c}$ ${\to}$ $BP$, $BV$ decays have been
  obtained with different approach in previous works, where
  the same value of coefficient $a_{1,2}$ is taken.
  The disagreement among previous works is largely originated
  from the different values of form factor.
  If the same value of form factors are used, the disparities
  on branching ratios of $a_{1}$-dominated $B_{c}$ ${\to}$
  $B_{d,s}P$, $B_{d,s}V$ decays will be greatly weakened.
  For example, if the same $F_{0}^{B_{c}{\to}B_{s}}$
  $=$ $1.0$ is fixed in the previous works,
  the branching ratio for $B_{c}$ ${\to}$ $B_{s}{\pi}$
  decays will all be about 10\%,
  which is consistent with our estimation within
  uncertainties and also agrees with the LHCb
  measurement \cite{1308.4544}.

  (2)
  From Table.\ref{tab05}, it can be seen that
  there are hierarchy between the branching ratios for
  $B_{c}$ ${\to}$ $BP$ and $B_{c}$ ${\to}$ $BV$
  decays with the same $B_{q}$ meson in the final state,
  for example,
    \begin{equation}
   {\cal B}r(B_{c}{\to}B_{q}{\pi})\ >\ {\cal B}r(B_{c}{\to}B_{q}{\rho})\
    >\ {\cal B}r(B_{c}{\to}B_{q}{\omega})
   \label{bqpi-bqrho},
   \end{equation}
   \begin{equation}
   {\cal B}r(B_{c}{\to}B_{q}K)\ >\ {\cal B}r(B_{c}{\to}B_{q}K^{\ast})
   \label{bqk-bqkstar},
   \end{equation}
  which differ from the previous prediction (see Table.\ref{tab01}).
  Two factors had a decisive influence on the above relations.
  One is kinematic factor. The phase space for $B_{c}$ ${\to}$
  $BP$ decay is larger than that for $B_{c}$ ${\to}$ $BV$
  decay, besides the orbital angular momentum
  $L_{BP}$ $<$ $L_{BV}$.
  The other is the form factor $F_{1}^{B_{c}{\to}B}(q^{2})$.
  For example, in the previous work \cite{prd.86.094028},
  the $F_{1}^{B_{c}{\to}B}(q^{2})$ goes up along with
  the growth of $q^{2}$,
  while in this paper, the shape line of
  $F_{1}^{B_{c}{\to}B}(q^{2})$ goes down in large
  $q^{2}$ region.
  The hierarchy between the branching ratios for
  $B_{c}$ ${\to}$ $BP$ and $B_{c}$ ${\to}$ $BV$
  decays can be serve as a standard to distinguish
  different approach, to check the practicality of
  the pQCD approach.

  (3) As noticed in \cite{prd.77.114004},
  the contributions of both penguin and annihilation
  to the branching ratios are very small for $B_{c}$
  ${\to}$ $BP$, $BV$ decay, because they are seriously
  suppressed by the CKM factors.
   \begin{center}
   \begin{tabular}{l|c|c} \hline \hline
   \multicolumn{1}{c|}{tree} & penguin & annihilation \\ \hline
       $V_{ud}V_{cs}^{\ast}$ ${\sim}$ $1$,~~~~
       $V_{us}V_{cs}^{\ast}$ ${\sim}$ $+{\lambda}$
     & $V_{ud}V_{cd}^{\ast}$ $+$ $V_{us}V_{cs}^{\ast}$ ${\sim}$ ${\lambda}^{5}$
     & $V_{cb}V_{ub}^{\ast}$ ${\sim}$ ${\lambda}^{5}$  \\
       $V_{us}V_{cd}^{\ast}$ ${\sim}$ ${\lambda}^{2}$,~~~
       $V_{ud}V_{cd}^{\ast}$ ${\sim}$ $-{\lambda}$ & &\\ \hline \hline
   \end{tabular}
   \end{center}

 There are large destructive interferences between the CKM
 factor $V_{ud}V_{cd}^{\ast}$ $\sim$ $-{\lambda}$ associated
 to decay amplitude ${\cal A}(B_{c}{\to}B_{u}{\eta}_{q})$
 and $V_{us}V_{cs}^{\ast}$ ${\sim}$ $+{\lambda}$ related to
 decay amplitude ${\cal A}(B_{c}{\to}B_{u}{\eta}_{s})$.
 In addition, the annihilation contribution is proportional to
 the color-favored tree parameter $a_{1}$.
 Hence, a significant annihilation contribution appear in the
 $B_{c}$ ${\to}$ $B_{u}{\eta}^{({\prime})}$ decays.

 (4) As noticed in \cite{prd.77.114004},
 due to the parameter $a_{1,2}$ and the CKM factors,
 there is hierarchy of amplitudes among branching ratios
 for the $B_{c}$ ${\to}$ $BP$, $BV$ decays.
   \begin{center}
   \begin{tabular}{l|c|c|c} \hline \hline
    \multicolumn{1}{c|}{mode}
    & parameter
    & CKM factor
    & branching ratio \\ \hline
  $B_{c}$ ${\to}$ $B_{s}{\pi}$, $B_{s}{\rho}$
    & $a_{1}$
    & $V_{ud}V_{cs}^{\ast}$ ${\sim}$ $1$
    & ${\cal O}(10^{-2})$ \\
  $B_{c}$ ${\to}$ $B_{s}K^{(\ast)}$
    & $a_{1}$
    & $V_{us}V_{cs}^{\ast}$ ${\sim}$ ${\lambda}$
    & $10^{-3}$ ${\sim}$ $10^{-5}$ \\ \hline
  $B_{c}$ ${\to}$ $B_{d}{\pi}$, $B_{d}{\rho}$
    & $a_{1}$
    & $V_{ud}V_{cd}^{\ast}$ ${\sim}$ ${\lambda}$
    & ${\cal O}(10^{-3})$ \\
   $B_{c}$ ${\to}$ $B_{d}K^{(\ast)}$
    & $a_{1}$
    & $V_{us}V_{cd}^{\ast}$ ${\sim}$ ${\lambda}^{2}$
    & $10^{-4}$ ${\sim}$ $10^{-5}$ \\ \hline
   $B_{c}^{+}$ ${\to}$ $B_{u}^{+}\overline{K}^{(\ast)}$
    & $a_{2}$
    & $V_{ud}V_{cs}^{\ast}$ ${\sim}$ $1$
    & $10^{-3}$ ${\sim}$ $10^{-4}$ \\
   $B_{c}$ ${\to}$ $B_{u}{\pi}$, $B_{u}{\rho}$, $B_{u}{\omega}$
    & $a_{2}$
    & $V_{ud}V_{cd}^{\ast}$ ${\sim}$ ${\lambda}$
    & ${\cal O}(10^{-5})$ \\
   $B_{c}^{+}$ ${\to}$ $B_{u}^{+}K^{(\ast)}$
    & $a_{2}$
    & $V_{us}V_{cd}^{\ast}$ ${\sim}$ ${\lambda}^{2}$
    & $10^{-6}$ ${\sim}$ $10^{-7}$ \\ \hline \hline
  \end{tabular}
  \end{center}
  Here, the branching ratios for the $B_{c}$ ${\to}$ $B_{u}P$,
  $B_{u}V$ decays are larger than those listed in
  \cite{prd.77.114004}. There are two reasons. One is that
  the decay amplitudes for the $B_{c}$ ${\to}$ $B_{u}P$,
  $B_{u}V$ decays is proportional to parameter $a_{2}$, and
  the value of $a_{2}$ in the ${\alpha}_{s}/{\pi}$ $\geq$ $0.15$
  region is much larger than $a_{2}(m_{c})$ used in
  \cite{prd.77.114004}.
  The other is that the nonfactorizable contributions
  $M_{cd,1}^{P,V}$ are
  always multiplied by the large Wilson coefficient $C_{1}$
  [see Eq.(\ref{bukb}-\ref{buetas})], which can largely
  enhance the branching ratios of color-suppressed tree
  $B_{c}$ ${\to}$ $B_{u}P$, $B_{u}V$ decays.

  (5) There are large uncertainties to the branching ratios
  from the shape parameter ${\omega}_{B_{q}}$ and the scale.
  Our numerical results are very rough.
  Despite this, we still get some information about the
  $B_{c}$ ${\to}$ $BP$, $BV$ decays. For example,
  the branching ratios for $B_{c}$ ${\to}$ $B_{d,s}{\pi}$,
  $B_{d,s}{\rho}$, $B_{s}K$ are large, these decay modes
  could clearly be measured by the running LHCb soon.

  \section{Summary}
  \label{sec4}
  In prospects of the potential $B_{c}$ meson at the LHCb
  experiments, accurate and thorough studies of the
  $B_{c}$ physics will be accessible very soon.
  In this paper, we calculated the $B_{c}$ ${\to}$
  $B_{u,d,s}$ transition form factors defined in vector
  and axial vector currents using the pQCD approach.
  We find that with appropriate scale, keeping
  the quark transverse momentum and introducing the
  Sudakov factors to modify the endpoint behavior,
  about 90\% contributions to the form factors comes
  form the ${\alpha}_{s}/{\pi}$ $<$ $0.3$ region.
  We studied the seventeen exclusive two-body hadronic
  $B_{c}$ ${\to}$ $B_{q}P$, $B_{q}V$ decays.
  It is shown that the contributions of penguin and
  annihilation to branching ratios are very small,
  because they relative to the tree contribution
  are highly suppressed by the CKM factors.
  The branching ratios for $B_{c}$ ${\to}$ $B_{d,s}{\pi}$,
  $B_{d,s}{\rho}$, $B_{s}K$ are large and could be
  easily measured by the running LHCb in the near
  future.

  \section*{Acknowledgments}
  This work is supported by {\em National Natural Science
  Foundation of China} under Grant Nos. 11147008, 11275057,
  11105043 and U1232101). Q. Chang is also supported by
  {\em Research Fund for the Doctoral Program of Higher
  Education of China under Grant No. 20114104120002,
  Foundation for the Author of National Excellent Doctoral
  Dissertation of P. R. China under Grant No. 201317},
  and {\em Program for Science and Technology Innovation
  Talents in Universities of Henan Province}.
  We would like to thank Prof. Hsiangnan Li,
  Prof. Caidian L\"{u}, Prof. Zhenjun Xiao for their helpful
  discussion, good counsel and encouragement.

  \begin{appendix}
  \section{distribution amplitudes of $B$ meson}
  \label{app01}
 For the heavy-light $B_{q}$ meson ($q$ $=$ $u$, $d$, $s$),
 we will adopt the Gaussian type distribution amplitudes
 proposed in \cite{Phys.Rev.D63.054008},
  \begin{equation}
 {\phi}_{B_{q}}(x,b)=
  Nx^{2}\bar{x}^{2}\,
 {\exp}\Big\{-\frac{1}{2}\Big(\frac{x\,m_{B_{q}}}{\omega}\Big)^{2}
             -\frac{1}{2}{\omega}^{2}b^{2}\Big\}
  \label{gaussian},
  \end{equation}
  where $N$ is the normalization constant.
  The shape of the distribution amplitude ${\phi}_{B_{q}}(x,0)$
  is displayed in Fig. \ref{fig06}.
  It is easy to see that the large value of shape parameter
  ${\omega}$ gives a large momentum fraction to the light
  spectator quark in $B_{q}$ meson.
  Because the mass of $s$ quark is heavier than that of $u,d$
  quark, it is assumed that the momentum fraction of the
  spectator quark $s$ in $B_{s}$ meson should be larger than
  that of the spectator quark $u,d$ in $B_{u,d}$ meson.
  In our calculation, we will use
  ${\omega}$ $=$ $0.45{\pm}0.05$ GeV for $B_{u,d}$ meson and
  ${\omega}$ $=$ $0.55{\pm}0.05$ GeV for $B_{s}$ meson.

  Due to the fact $m_{B_{c}}$ ${\approx}$ $m_{b}$ $+$ $m_{c}$
  the $B_{c}$ meson can be approximated as a non-relativistic
  bound state of two heavy quark $b$ and $c$.
  Its wave function is approximately the solution of the
  Schr\"{o}dinger equation with the harmonic oscillator potential.
  For the ground pseudoscalar $B_{c}$ meson, the corresponding
  radial wave function is
   \begin{equation}
  {\psi}_{nL}(r)
 ={\psi}_{1S}(r)\
  {\propto}\
  {\exp}(-{\alpha}^{2}r^{2}/2)
   \label{schrodinger01},
   \end{equation}
  where ${\alpha}^{2}$ $=$ ${\mu}{\omega}$,
  the reduced mass ${\mu}$ $=$ $m_{b}m_{c}/(m_{b}+m_{c})$
  and the quantum of energy ${\omega}$ ${\approx}$
  $0.50{\pm}0.05$ GeV \cite{omega}.

  Applying the Fourier transform, one can get the
  representation of wave function in momentum space
   \begin{equation}
  {\psi}_{1S}(\vec{k})\
  {\sim}\
  {\int}{\bf d}\vec{r}\
  {\psi}_{1S}(r)
  e^{-i{\vec{k}{\cdot}\vec{r}}}\
  {\propto}\
  {\exp}(-k^{2}/2{\alpha}^{2})
   \label{schrodinger02}.
   \end{equation}
  Then adopting the connection \cite{Phys.Rev.D22.2157}
  between the equal-time prescription in the rest frame
  and the light-cone dynamics, i.e., assuming that the
  constituent quarks $b$ and $c$ are on-shell and their
  light-cone momentum fraction are $x_{b}$ and $x_{c}$,
  with $x_{b}$ $+$ $x_{c}$ $=$ $1$, one can get the
  light-cone wave function for $B_{c}$ meson,
   \begin{equation}
  {\psi}_{B_{c}}(x_{i},\vec{k}_{\perp})\
  {\propto}\
  {\exp}\Big\{-\frac{1}{8{\alpha}^{2}}
   \Big(\frac{\vec{k}_{\perp}^{2}+m_{c}^{2}}{x_{c}}
      + \frac{\vec{k}_{\perp}^{2}+m_{b}^{2}}{x_{b}}
   \Big)\Big\}
   \label{schrodinger03}.
   \end{equation}
  The distribution amplitudes of $B_{c}$ meson is
   \begin{eqnarray}
  {\phi}_{B_{c}}(x_{i})&=&
  {\int}{\bf d}\vec{k}_{\perp}
  {\psi}_{B_{c}}(x_{i},\vec{k}_{\perp})
   \nonumber \\ &=& N
   \frac{x_{b}x_{c}}{x_{b}+x_{c}}
  {\exp}\Big\{-\frac{1}{8{\alpha}^{2}}
   \Big(\frac{m_{c}^{2}}{x_{c}}
      + \frac{m_{b}^{2}}{x_{b}}
   \Big)\Big\}
   \label{schrodinger04},
   \end{eqnarray}
  where $N$ is the normalization constant and
  the normalization condition is
   \begin{equation}
  {\int}{\bf d}x\,{\phi}_{B_{c}}(x)=1
   \label{schrodinger05}.
   \end{equation}
  In our calculation, $x$ $=$ $x_{c}$ and
  $\bar{x}$ $=$ $x_{b}$ $=$ $1$ $-$ $x$, so
  we have
   \begin{equation}
  {\phi}_{B_{c}}(x)=Nx\bar{x}\
  {\exp}\Big\{-\frac{1}{8{\alpha}^{2}}
   \Big(\frac{m_{c}^{2}}{x}
      + \frac{m_{b}^{2}}{\bar{x}}
   \Big)\Big\}
   \label{schrodinger06}.
   \end{equation}
  The shape of the distribution amplitude of $B_{c}$ meson
  is displayed in Fig. \ref{fig07}. It is easy to see that
  the maximum position is near $m_{c}/(m_{b}+m_{c})$ and
  that the small value of parameter ${\omega}$ gives
  a narrow shape. In our calculation,
  we will use ${\omega}$ $=$ $0.50{\pm}0.05$ GeV for $B_{c}$ meson.
  \section{distribution amplitudes of light mesons}
  \label{app02}
 The twist-2 quark-antiquark distribution amplitudes of
 light pseudoscalar and longitudinally polarized vector
 meson are expressed as
 \cite{JHEP.01.010.1999,prd.65.014007,Nucl.Phys.B529.323},
   \begin{eqnarray}
  {\phi}_{P}^{a}(x)
  &=&
  6x\bar{x}\sum\limits_{n}
  a_{n}C_{n}^{3/2}({\xi})
   \label{wfpa}, \\
  {\phi}_{V}(x)
  &=&
  6x\bar{x}\sum\limits_{n}
  a_{n}^{\parallel}C_{n}^{3/2}({\xi})
   \label{wfvv},
   \end{eqnarray}
  where $C_{n}^{3/2}({\xi})$ is the Gegenbauer polynomial,
  and ${\xi}$ $=$ $x$ $-$ $\bar{x}$ $=$ $2x$ $-$ $1$.
  The Gegenbauer moments $a_{0}$ $=$ $1$ and
  $a_{0}^{\parallel}$ $=$ $1$ due to the normalization
  condition
   \begin{equation}
  {\int}_{0}^{1}{\bf d}x\,{\phi}_{P}^{a}(x)=
  {\int}_{0}^{1}{\bf d}x\,{\phi}_{V}(x)=1
   \label{norm-pv}.
   \end{equation}

  The two-particle twist-3 distribution amplitudes of
  pseudoscalar meson have the expansion in the terms
  of the Gegenbauer polynomials
  \cite{JHEP.01.010.1999,prd.65.014007},
   \begin{eqnarray}
  {\phi}_{P}^{p}(x)
  &=& 1
   +\Big(30{\eta}_{3}-\frac{5}{2}{\rho}_{P}^{2}\Big)C_{2}^{1/2}({\xi})
   -\Big(3{\eta}_{3}{\omega}_{3}+\frac{27}{20}{\rho}_{P}^{2}
   + \frac{81}{10}{\rho}_{P}^{2}a_{2}\Big)C_{4}^{1/2}({\xi})
   \label{wfpp}, \\
  {\phi}_{P}^{t}(x)
  &=&
    C_{1}^{1/2}(-{\xi})
  +6\Big(5{\eta}_{3}
  -\frac{1}{2}{\eta}_{3}{\omega}_{3}
  -\frac{7}{20}{\rho}_{P}^{2}
  -\frac{3}{5}{\rho}_{P}^{2}a_{2}\Big)
  C_{3}^{1/2}(-{\xi})
   \label{wfpt}.
   \end{eqnarray}

  The expressions of the two-particle twist-3 distribution
  amplitudes of the longitudinally polarized vector meson
  are \cite{prd.65.014007,Nucl.Phys.B529.323}
   \begin{eqnarray}
  {\phi}_{V}^{t}(x)
  &=&3{\xi}^{2}
   \label{wfvt}, \\
  {\phi}_{V}^{s}(x)
  &=&-3{\xi}
   \label{wfvs}.
   \end{eqnarray}

  In the mesonic distribution amplitudes, the Gegenbauer
  polynomials are
   \begin{eqnarray}
   C_{1}^{1/2}(x)&=&x
   \label{c11}, \\
   C_{2}^{1/2}(x)&=&
   \frac{1}{2}(3x^{2}-1)
   \label{c21}, \\
   C_{3}^{1/2}(x)&=&
   \frac{1}{2}(5x^{3}-3x)
   \label{c31}, \\
   C_{4}^{1/2}(x)&=&
   \frac{1}{8}(35x^{4}-30x^{2}+3)
   \label{c41}, \\
   C_{1}^{3/2}(x)&=&3x
   \label{c13}, \\
   C_{2}^{3/2}(x)&=&
   \frac{3}{2}(5x^{2}-1)
   \label{c23}, \\
   C_{3}^{3/2}(x)&=&
   \frac{5}{2}(7x^{3}-3x)
   \label{c33}, \\
   C_{4}^{3/2}(x)&=&
   \frac{15}{8}(21x^{4}-14x^{2}+1)
   \label{c43}.
   \end{eqnarray}
  \section{formula of decay amplitude}
  \label{app03}
  The decay amplitudes can be expressed in terms of the following
  building block:
   \begin{eqnarray}
   C_{P}&=&
   \frac{C_{F}{\pi}}{N_{c}}f_{B_{c}}f_{B_{q}}f_{P}
   \label{cpseudoscalar}, \\
   C_{V}&=&
   \frac{C_{F}{\pi}}{N_{c}}f_{B_{c}}f_{B_{q}}f_{V}
   \label{cvector},
   \end{eqnarray}
   \begin{eqnarray}
  iM^{P}_{a,1}&=& C_{P}
  {\int}_{0}^{1}{\bf d}x_{1}{\bf d}x_{2}
  {\int}_{0}^{\infty}{\bf d}b_{1}{\bf d}b_{2}
  {\phi}_{B_{c}}(x_{1})
  {\phi}_{B_{q}}(x_{2},b_{2})H_{a}
   \nonumber \\ &{\times}&
   \{
   (m_{c}-x_{2}m_{2})(m_{2}u-2m_{1}s)
  +(x_{2}s+m_{3}^{2})(t-4m_{1}m_{2})
   \}
   \label{mpa1}, \\
   iM^{P}_{a,2}&=&-iM^{P}_{a,2}
   \label{mpa2}, \\
  iM^{P}_{a,3}&=& C_{P}
  {\int}_{0}^{1}{\bf d}x_{1}{\bf d}x_{2}
  {\int}_{0}^{\infty}{\bf d}b_{1}{\bf d}b_{2}
  {\phi}_{B_{c}}(x_{1})
  {\phi}_{B_{q}}(x_{2},b_{2})H_{a}
   \nonumber \\ &{\times}&
   2{\mu}_{P} \{
   (m_{c}+x_{2}m_{2})(t-4m_{1}m_{2})
  +(m_{2}u-2m_{1}s) \}
   \label{mpa3},
   \end{eqnarray}
   \begin{eqnarray}
  iM^{P}_{b,1}&=& C_{P}
  {\int}_{0}^{1}{\bf d}x_{1}{\bf d}x_{2}
  {\int}_{0}^{\infty}{\bf d}b_{1}{\bf d}b_{2}
  {\phi}_{B_{c}}(x_{1})
  {\phi}_{B_{q}}(x_{2},b_{2})H_{b}
   \nonumber \\ &{\times}&
   \{ x_{1}m_{1}(2m_{2}u-m_{1}s)
    +(x_{1}u-m_{3}^{2})(t-4m_{1}m_{2})
   \}
   \label{mpb1}, \\
   iM^{P}_{b,2}&=&-iM^{P}_{b,2}
   \label{mpb2}, \\
  iM^{P}_{b,3}&=& C_{P}
  {\int}_{0}^{1}{\bf d}x_{1}{\bf d}x_{2}
  {\int}_{0}^{\infty}{\bf d}b_{1}{\bf d}b_{2}
  {\phi}_{B_{c}}(x_{1})
  {\phi}_{B_{q}}(x_{2},b_{2})H_{b}
   \nonumber \\ &{\times}&
 2{\mu}_{P} \{
   x_{1}m_{1}(t-4m_{1}m_{2})
  +(2m_{2}u-m_{1}s) \}
   \label{mpb3},
   \end{eqnarray}
   \begin{eqnarray}
  iM^{P}_{c,1}&=& C_{P}
  {\int}_{0}^{1}{\bf d}x_{1}{\bf d}x_{2}{\bf d}x_{3}
  {\int}_{0}^{\infty}{\bf d}b_{2}{\bf d}b_{3}
  {\phi}_{B_{c}}(x_{1})
  {\phi}_{B_{q}}(x_{2},b_{2})
  {\phi}_{P}^{a}(x_{3})H_{c}
   \nonumber \\ &{\times}&
   \{ s\,t(x_{1}-x_{2})
     +s\,m_{1}m_{2}(x_{2}-x_{3})
     +u(s-m_{1}m_{2})(x_{1}-x_{3})
   \}
   \label{mpc1}, \\
   iM^{P}_{c,2}&=& C_{P}
  {\int}_{0}^{1}{\bf d}x_{1}{\bf d}x_{2}{\bf d}x_{3}
  {\int}_{0}^{\infty}{\bf d}b_{2}{\bf d}b_{3}
  {\phi}_{B_{c}}(x_{1})
  {\phi}_{B_{q}}(x_{2},b_{2})
  {\phi}_{P}^{a}(x_{3})H_{c}
   \nonumber \\ &{\times}&
   \{
   u\,t(x_{2}-x_{1})
  +u\,m_{1}m_{2}(x_{1}-x_{3})
  +s(u+m_{1}m_{2})(x_{3}-x_{2})
   \}
   \label{mpc2}, \\
  iM^{P}_{c,3}&=& C_{P}
  {\int}_{0}^{1}{\bf d}x_{1}{\bf d}x_{2}{\bf d}x_{3}
  {\int}_{0}^{\infty}{\bf d}b_{2}{\bf d}b_{3}
  {\phi}_{B_{c}}(x_{1})
  {\phi}_{B_{q}}(x_{2},b_{2})
  {\mu}_{P} H_{c}
   \nonumber \\ &{\times}&
   \{
  {\phi}_{P}^{p}(x_{3}) [
   u\,m_{2}(x_{1}-x_{3})
  +s\,m_{1}(x_{2}-x_{3})
  +t(m_{1}+m_{2})(x_{1}-x_{2})]
   \nonumber \\ &&+
  {\phi}_{P}^{t}(x_{3})\,2m_{1}p\,
   [ m_{1}(x_{1}-x_{3})
   + m_{2}(x_{2}-x_{3}) ]
   \}
   \label{mpc3},
   \end{eqnarray}
   \begin{eqnarray}
  iM^{P}_{d,1}&=& C_{P}
  {\int}_{0}^{1}{\bf d}x_{1}{\bf d}x_{2}{\bf d}x_{3}
  {\int}_{0}^{\infty}{\bf d}b_{2}{\bf d}
  {\phi}_{B_{c}}(x_{1})
  {\phi}_{B_{q}}(x_{2},b_{2})
  {\phi}_{P}^{a}(x_{3}) H_{d}
   \nonumber \\ &{\times}&
   \{
   u\,t(x_{2}-x_{1})
  +u\,m_{1}m_{2}(x_{1}-\bar{x}_{3})
  +s(u+m_{1}m_{2})(\bar{x}_{3}-x_{2})
   \}
   \label{mpd1}, \\
   iM^{P}_{d,2}&=& C_{P}
  {\int}_{0}^{1}{\bf d}x_{1}{\bf d}x_{2}{\bf d}x_{3}
  {\int}_{0}^{\infty}{\bf d}b_{2}{\bf d}
  {\phi}_{B_{c}}(x_{1})
  {\phi}_{B_{q}}(x_{2},b_{2})
  {\phi}_{P}^{a}(x_{3}) H_{d}
   \nonumber \\ &{\times}&
   \{
   s\,t(x_{1}-x_{2})
  +s\,m_{1}m_{2}(x_{2}-\bar{x}_{3})
  +u(s-m_{1}m_{2})(x_{1}-\bar{x}_{3})
   \}
   \label{mpd2}, \\
  iM^{P}_{d,3}&=& C_{P}
  {\int}_{0}^{1}{\bf d}x_{1}{\bf d}x_{2}{\bf d}x_{3}
  {\int}_{0}^{\infty}{\bf d}b_{2}{\bf d}
  {\phi}_{B_{c}}(x_{1})
  {\phi}_{B_{q}}(x_{2},b_{2})
  {\mu}_{P} H_{d}
   \nonumber \\ &{\times}&
   \{
  {\phi}_{P}^{p}(x_{3}) [
   u\,m_{2}(\bar{x}_{3}-x_{1})
  +s\,m_{1}(\bar{x}_{3}-x_{2})
  +t(m_{1}+m_{2})(x_{2}-x_{1}) ]
   \nonumber \\ & &+
  {\phi}_{P}^{t}(x_{3})\,2m_{1}p\, [
    m_{1}(x_{1}-\bar{x}_{3})
   +m_{2}(x_{2}-\bar{x}_{3}) ]
   \}
   \label{mpd3},
   \end{eqnarray}
   \begin{eqnarray}
  iM^{P}_{e,1}&=& C_{P}
  {\int}_{0}^{1}{\bf d}x_{2}{\bf d}x_{3}
  {\int}_{0}^{\infty}{\bf d}b_{2}{\bf d}b_{3}
  {\phi}_{B_{q}}(x_{2},b_{2}) H_{e}
   \nonumber \\ &{\times}&
   \{
  {\phi}_{P}^{a}(x_{3})
   [x_{2}m_{1}^{2}s+\bar{x}_{2}m_{3}^{3}t)]
   +
  {\mu}_{P}{\phi}_{P}^{p}(x_{3})\,2m_{2}
   [x_{2}t+u]
   \}
   \label{mpe1},
   \end{eqnarray}
   \begin{eqnarray}
  iM^{P}_{f,1}&=& C_{P}
  {\int}_{0}^{1}{\bf d}x_{2}{\bf d}x_{3}
  {\int}_{0}^{\infty}{\bf d}b_{2}{\bf d}b_{3}
  {\phi}_{B_{q}}(x_{2},b_{2}) H_{f}
   \nonumber \\ &{\times}&
   \{
  {\phi}_{P}^{a}(x_{3}) [
    2m_{2}m_{b}u
   -\bar{x}_{3}m_{1}^{2}s
   -x_{3}m_{2}^{2}u ]
  \nonumber \\ &&+
 {\mu}_{P}{\phi}_{P}^{p}(x_{3})[
  m_{b}\,t-2m_{2}(t+\bar{x}_{3}u)]
  \nonumber \\ &&+
 {\mu}_{P}{\phi}_{P}^{t}(x_{3})\,2m_{1}p\,
  [m_{b}-2m_{2}x_{3}]
  \}
  \label{mpf1},
  \end{eqnarray}
   \begin{eqnarray}
  iM^{P}_{g,1}&=& C_{P}
  {\int}_{0}^{1}{\bf d}x_{1}{\bf d}x_{2}{\bf d}x_{3}
  {\int}_{0}^{\infty}{\bf d}b_{1}{\bf d}b_{2}
  {\phi}_{B_{c}}(x_{1})
  {\phi}_{B_{q}}(x_{2},b_{2})H_{g}
   \nonumber \\ &{\times}&
   \{
   {\phi}_{P}^{a}(x_{3}) [
   s\,t(\bar{x}_{2}-x_{3})
  +t\,u(x_{3}-x_{1})
  -m_{1}m_{b}s ]
   \nonumber \\ & &+
  {\mu}_{P}{\phi}_{P}^{p}(x_{3}) m_{2} [
   t(\bar{x}_{2}-x_{1})
  +u(x_{3}-x_{1})-4m_{1}m_{b} ]
   \nonumber \\ & &+
  {\mu}_{P}{\phi}_{P}^{t}(x_{3})\,
   2m_{1}m_{2}p\,
  (x_{3}-\bar{x}_{2})
   \}
   \label{mpg1},
   \end{eqnarray}
   \begin{eqnarray}
  iM^{P}_{h,1}&=& C_{P}
  {\int}_{0}^{1}{\bf d}x_{1}{\bf d}x_{2}{\bf d}x_{3}
  {\int}_{0}^{\infty}{\bf d}b_{1}{\bf d}b_{2}
  {\phi}_{B_{c}}(x_{1})
  {\phi}_{B_{q}}(x_{2},b_{2})H_{h}
   \nonumber \\ &{\times}&
   \{
  {\phi}_{P}^{a}(x_{3}) [
   s\,u(\bar{x}_{3}-x_{2})
  +t\,u(x_{2}-x_{1})
  +m_{1}m_{c}s  ]
   \nonumber \\ & &+
  {\mu}_{P}{\phi}_{P}^{p}(x_{3}) m_{2} [
   t(x_{2}-x_{1})
  +u(\bar{x}_{3}-x_{1})
  +4m_{1}m_{c}]
   \nonumber \\ & &+
  {\mu}_{P}{\phi}_{P}^{t}(x_{3})\,
  2m_{1}m_{2}p\,
  (x_{2}-\bar{x}_{3})
   \}
   \label{mph1},
   \end{eqnarray}
   \begin{eqnarray}
  M^{V}_{a,1}&=& C_{V}m_{1}p
  {\int}_{0}^{1}{\bf d}x_{1}{\bf d}x_{2}
  {\int}_{0}^{\infty}{\bf d}b_{1}{\bf d}b_{2}
  {\phi}_{B_{c}}(x_{1})
  {\phi}_{B_{q}}(x_{2},b_{2})H_{a}
   \nonumber \\ &{\times}&
   \{
   x_{2}(t+s-4m_{1}m_{2})
   -2m_{c}(2m_{1}-m_{2})
   +2m_{3}^{2} \}
   \label{mva1}, \\
  M^{V}_{a,2}&=&M^{V}_{a,1}
   \label{mva2}, \\
  M^{V}_{a,3}&=&0
   \label{mva3},
   \end{eqnarray}
   \begin{eqnarray}
  M^{V}_{b,1}&=& C_{V}m_{1}p
  {\int}_{0}^{1}{\bf d}x_{1}{\bf d}x_{2}
  {\int}_{0}^{\infty}{\bf d}b_{1}{\bf d}b_{2}
  {\phi}_{B_{c}}(x_{1})
  {\phi}_{B_{q}}(x_{2},b_{2})H_{b}
   \nonumber \\ &{\times}&
   \{ x_{1}(t-u-4m_{1}m_{2})+2m_{3}^{2} \}
   \label{mvb1}, \\
  M^{V}_{b,2}&=&M^{V}_{b,1}
   \label{mvb2}, \\
  M^{V}_{b,3}&=&0
   \label{mvb3},
   \end{eqnarray}
   \begin{eqnarray}
  M^{V}_{c,1}&=& C_{V}m_{1}p
  {\int}_{0}^{1}{\bf d}x_{1}{\bf d}x_{2}{\bf d}x_{3}
  {\int}_{0}^{\infty}{\bf d}b_{2}{\bf d}b_{3}
  {\phi}_{B_{c}}(x_{1})
  {\phi}_{B_{q}}(x_{2},b_{2})
   \nonumber \\ &{\times}&
   2{\phi}_{V}(x_{3})H_{c} \{
    (t-m_{1}m_{2})(x_{1}-x_{2})
   +u(x_{1}-x_{3}) \}
   \label{mvc1}, \\
   M^{V}_{c,2}&=& C_{V}m_{1}p
  {\int}_{0}^{1}{\bf d}x_{1}{\bf d}x_{2}{\bf d}x_{3}
  {\int}_{0}^{\infty}{\bf d}b_{2}{\bf d}b_{3}
  {\phi}_{B_{c}}(x_{1})
  {\phi}_{B_{q}}(x_{2},b_{2})
   \nonumber \\ &{\times}&
  2{\phi}_{V}(x_{3})H_{c} \{
    (t-m_{1}m_{2})(x_{1}-x_{2})
   +s(x_{2}-x_{3}) \}
   \label{mvc2}, \\
  M^{V}_{c,3}&=& C_{V}m_{3}\frac{f_{V}^{T}}{f_{V}}
  {\int}_{0}^{1}{\bf d}x_{1}{\bf d}x_{2}{\bf d}x_{3}
  {\int}_{0}^{\infty}{\bf d}b_{2}{\bf d}b_{3}
  {\phi}_{B_{c}}(x_{1})
  {\phi}_{B_{q}}(x_{2},b_{2})
   \nonumber \\ &{\times}&
   H_{c} \{
  {\phi}_{V}^{t}(x_{3})\,2m_{1}p\,[
    m_{1}(x_{3}-x_{1})
   +m_{2}(x_{3}-x_{2}) ]
   \nonumber \\ & &~+
  {\phi}_{V}^{s}(x_{3}) [
    m_{2}u(x_{3}-x_{1})
   +m_{1}s(x_{3}-x_{2})
   \nonumber \\ & &~~~~~~~~~~~~+
    t\,(m_{1}+m_{2})(x_{2}-x_{1})
    ] \}
   \label{mvc3},
   \end{eqnarray}
   \begin{eqnarray}
  M^{V}_{d,1}&=& C_{V}m_{1}p
   {\int}_{0}^{1}{\bf d}x_{1}{\bf d}x_{2}{\bf d}x_{3}
  {\int}_{0}^{\infty}{\bf d}b_{2}{\bf d}b_{3}
  {\phi}_{B_{c}}(x_{1})
  {\phi}_{B_{q}}(x_{2},b_{2})
   \nonumber \\ &{\times}&
   2{\phi}_{V}(x_{3})H_{d} \{
   (t-m_{1}m_{2})(x_{2}-x_{1})
   +s(\bar{x}_{3}-x_{2}) \}
   \label{mvd1}, \\
   M^{V}_{d,2}&=&C_{V}m_{1}p
   {\int}_{0}^{1}{\bf d}x_{1}{\bf d}x_{2}{\bf d}x_{3}
  {\int}_{0}^{\infty}{\bf d}b_{2}{\bf d}b_{3}
  {\phi}_{B_{c}}(x_{1})
  {\phi}_{B_{q}}(x_{2},b_{2})
   \nonumber \\ &{\times}&
   2{\phi}_{V}(x_{3})H_{d} \{
   (t-m_{1}m_{2})(x_{2}-x_{1})
   +u(\bar{x}_{3}-x_{1}) \}
   \label{mvd2}, \\
  M^{V}_{d,3}&=& C_{V}m_{3} \frac{f_{V}^{T}}{f_{V}}
   {\int}_{0}^{1}{\bf d}x_{1}{\bf d}x_{2}{\bf d}x_{3}
  {\int}_{0}^{\infty}{\bf d}b_{2}{\bf d}b_{3}
  {\phi}_{B_{c}}(x_{1})
  {\phi}_{B_{q}}(x_{2},b_{2})
   \nonumber \\ &{\times}&
   H_{d} \{
  {\phi}_{V}^{t}(x_{3})\,2m_{1}p\,
   [ m_{1}(\bar{x}_{3}-x_{1})
   + m_{2}(\bar{x}_{3}-x_{2}) ]
   \nonumber \\ & &~+
  {\phi}_{V}^{s}(x_{3}) [
   m_{2}u(x_{1}-\bar{x}_{3})
  +m_{1}s(x_{2}-\bar{x}_{3})
   \nonumber \\ & &~~~~~~~~~~~~+
   t\,(m_{1}+m_{2})(x_{1}-x_{2})
    ] \}
   \label{mvd3},
   \end{eqnarray}
   \begin{eqnarray}
  M^{V}_{e,1}&=& C_{V}
  {\int}_{0}^{1}{\bf d}x_{2}{\bf d}x_{3}
  {\int}_{0}^{\infty}{\bf d}b_{2}{\bf d}b_{3}
  {\phi}_{B_{q}}(x_{2},b_{2})H_{e}
   \nonumber \\ &{\times}&
   \big\{
  {\phi}_{V}(x_{3})\,m_{1}p\,
   [x_{2}(s+t)+2m_{3}^{2}]
   \nonumber \\ & &-
   {\phi}_{V}^{s}(x_{3})\,
   2m_{2}m_{3}\frac{f_{V}^{T}}{f_{V}}
   (x_{2}t+u) \big\}
   \label{mve1},
   \end{eqnarray}
   \begin{eqnarray}
  M^{V}_{f,1}&=& C_{V}
  {\int}_{0}^{1}{\bf d}x_{2}{\bf d}x_{3}
  {\int}_{0}^{\infty}{\bf d}b_{2}{\bf d}b_{3}
  {\phi}_{B_{q}}(x_{2},b_{2})H_{f}
   \nonumber \\ &{\times}&
   \{ {\phi}_{V}(x_{3})\,m_{1}p\,
   [x_{3}(s+u)+4m_{2}m_{b}-2m_{1}^{2} ]
   \nonumber \\ & &+
   m_{3}\frac{f_{V}^{T}}{f_{V}}
   {\phi}_{V}^{t}(x_{3})\,
   2m_{1}p\,(2m_{2}x_{3}-m_{b})
    \nonumber \\ & &+
   m_{3}\frac{f_{V}^{T}}{f_{V}}
   {\phi}_{V}^{s}(x_{3})
   [ t(2m_{2}-m_{b})+2m_{2}u\bar{x}_{3} ] \}
   \label{mvf1},
   \end{eqnarray}
   \begin{eqnarray}
  M^{V}_{g,1}&=& C_{V}
  {\int}_{0}^{1}{\bf d}x_{1}{\bf d}x_{2}{\bf d}x_{3}
  {\int}_{0}^{\infty}{\bf d}b_{1}{\bf d}b_{2}
  {\phi}_{B_{c}}(x_{1})
  {\phi}_{B_{q}}(x_{2},b_{2})H_{g}
   \nonumber \\ &{\times}&
   \{
  {\phi}_{V}(x_{3})\,2m_{1}p\,[
   t\,(\bar{x}_{2}-x_{1})-m_{1}m_{b} ]
   \nonumber \\ & &+
   m_{2}m_{3}\frac{f_{V}^{T}}{f_{V}}
  {\phi}_{V}^{t}(x_{3})\,2m_{1}p\,
  (\bar{x}_{2}-x_{3})
   \nonumber \\ & &+
   m_{2}m_{3}\frac{f_{V}^{T}}{f_{V}}
  {\phi}_{V}^{s}(x_{3}) [
   t\,(x_{1}-\bar{x}_{2})+u\,(x_{1}-x_{3})
   +4m_{1}m_{b} ] \}
   \label{mvg1},
   \end{eqnarray}
   \begin{eqnarray}
  M^{V}_{h,1}&=& C_{V}
  {\int}_{0}^{1}{\bf d}x_{1}{\bf d}x_{2}{\bf d}x_{3}
  {\int}_{0}^{\infty}{\bf d}b_{1}{\bf d}b_{2}
  {\phi}_{B_{c}}(x_{1})
  {\phi}_{B_{q}}(x_{2},b_{2})H_{h}
   \nonumber \\ &{\times}&
   \{ {\phi}_{V} (x_{3})\,2m_{1}p\,[
   t\,(x_{2}-x_{1})+s\,(\bar{x}_{3}-x_{2})
   +m_{1}m_{c}]
   \nonumber \\ & &+
   m_{2}m_{3}\frac{f_{V}^{T}}{f_{V}}
  {\phi}_{V}^{t}(x_{3})\,2m_{1}p\,
  (\bar{x}_{3}-x_{2})
   \nonumber \\ & &+
   m_{2}m_{3}\frac{f_{V}^{T}}{f_{V}}
  {\phi}_{V}^{s}(x_{3}) [
   t\,(x_{1}-x_{2})+u\,(x_{1}-\bar{x}_{3})
   -4m_{1}m_{c} ] \}
   \label{mvh1}.
   \end{eqnarray}

  The function $H_{i}$ are defined as
   \begin{eqnarray}
   H_{a} &=&b_{1}b_{2}
  e^{-S_{1}(t_{a})-S_{2}(t_{a})}
  {\alpha}_{s}(t_{a})
   K_{0}(\sqrt{{\alpha}_{e}}b_{1})
   \nonumber \\ &{\times}&
   \Big\{
  {\theta}(b_{1}-b_{2})
    K_{0}(\sqrt{{\beta}_{a}}b_{1})
    I_{0}(\sqrt{{\beta}_{a}}b_{2})
  +\big(b_{1}{\leftrightarrow}b_{2}\big)
   \Big\}
   \label{harda}, \\
   H_{b} &=&b_{1}b_{2}
  e^{-S_{1}(t_{b})-S_{2}(t_{b})}
  {\alpha}_{s}(t_{b})
   K_{0}(\sqrt{{\alpha}_{e}}b_{2})
   \nonumber \\ &{\times}&
   \Big\{
  {\theta}(b_{1}-b_{2})
   K_{0}(\sqrt{{\beta}_{b}}b_{1})
   I_{0}(\sqrt{{\beta}_{b}}b_{2})
  +\big(b_{1}{\leftrightarrow}b_{2}\big)
   \Big\}
   \label{hardb}, \\
   H_{i=c,d} &=&b_{2}b_{3}
  e^{-S_{1}(t_{i})-S_{2}(t_{i})-S_{3}(t_{i})}
  {\alpha}_{s}(t_{i})
   K_{0}(\sqrt{{\beta}_{i}}{b}_{3})
   \nonumber \\ &{\times}&
   \Big\{
  {\theta}(b_{2}-b_{3})
   K_{0}(\sqrt{{\alpha}_{e}}b_{2})
   I_{0}(\sqrt{{\alpha}_{e}}b_{3})
   +(b_{2}{\leftrightarrow}b_{3})
   \Big\}_{b_{1}=b_{2}}
   \label{hardcd}, \\
   H_{e} &=& b_{2}b_{3}
  e^{-S_{2}(t_{e})-S_{3}(t_{e})}
  {\alpha}_{s}(t_{e})
   K_{0}(\sqrt{-{\alpha}_{a}}b_{3})
   \nonumber \\ &{\times}&
   \Big\{
  {\theta}(b_{2}-b_{3})
   K_{0}(\sqrt{-{\beta}_{e}}b_{2})
   I_{0}(\sqrt{-{\beta}_{e}}b_{3})
   +(b_{2}{\leftrightarrow}b_{3})
   \Big\}
   \label{harde}, \\
   H_{f} &=& b_{2}b_{3}
  e^{-S_{2}(t_{f})-S_{3}(t_{f})}
  {\alpha}_{s}(t_{f})
   K_{0}(\sqrt{-{\alpha}_{a}}b_{2})
   \nonumber \\ &{\times}&
   \Big\{
  {\theta}(b_{2}-b_{3})
   K_{0}(\sqrt{{\beta}_{f}}b_{2})
   I_{0}(\sqrt{{\beta}_{f}}b_{3})
   +(b_{2}{\leftrightarrow}b_{3})
   \Big\}
   \label{hardf}, \\
   H_{i=g,h} &=& b_{1}b_{2}
  e^{-S_{1}(t_{i})-S_{2}(t_{i})-S_{3}(t_{i})}
  {\alpha}_{s}(t_{i})
   K_{0}(\sqrt{{\beta}_{i}}b_{1})
   \nonumber \\ &{\times}&
   \Big\{
  {\theta}(b_{1}-b_{2})
   K_{0}(\sqrt{-{\alpha}_{a}}b_{1})
   I_{0}(\sqrt{-{\alpha}_{a}}b_{2})
  +\big(b_{1}{\leftrightarrow}b_{2}\big)
   \Big\}_{b_{2}=b_{3}}
   \label{hardg}.
   \end{eqnarray}

  The exponent of the Sudakov factor $e^{-S}$ is given by
   \begin{eqnarray}
   S_{1}(t)&=&
   s(x_{1},b_{1},\frac{m_{1}}{\sqrt{2}})
   + \frac{5}{3}
    {\int}_{1/b_{1}}^{t}
     \frac{{\bf d}{\mu}}{\mu}
    {\gamma}_{q}({\mu})
   \label{suda01}, \\
   S_{2}(t)&=&
   s(x_{2},b_{2},q_{2}^{+})
   + \frac{5}{3}
    {\int}_{1/b_{2}}^{t}
     \frac{{\bf d}{\mu}}{\mu}
    {\gamma}_{q}({\mu})
   \label{suda02}, \\
   S_{3}(t)&=&
   s(x_{3},b_{3},q_{3}^{+})
  +s(\bar{x}_{3},b_{3},q_{3}^{+})
  +2{\int}_{1/b_{3}}^{t}
     \frac{{\bf d}{\mu}}{\mu}
    {\gamma}_{q}({\mu})
   \label{suda03},
   \end{eqnarray}
  where the function $s(x,b,Q)$ are defined in Appendix
  of Ref.\cite{Phys.Rev.D52.3958}.
  ${\gamma}_{q}$ $=$ $-{\alpha}_{s}/{\pi}$ is the
  quark anomalous dimension.

 The hard scale $t_{i}$ is chosen as the maximum of the
 virtuality of the internal quark and gluon, including
 $1/b$ (where $b$ is the transverse separation) i.e.,
   \begin{eqnarray}
   t_{i=a,b}&=&
  {\max}(
   \sqrt{{\alpha}_{e}},
   \sqrt{{\vert}{\beta}_{i}{\vert}},
  1/b_{1},1/b_{2})
   \label{tab}, \\
   t_{i=c,d}&=&
  {\max}(
   \sqrt{{\alpha}_{e}},
   \sqrt{{\vert}{\beta}_{i}{\vert}},
   1/b_{2},1/b_{3})
   \label{tcd}, \\
   t_{i=e,f}&=&
  {\max}(
   \sqrt{{\alpha}_{a}},
   \sqrt{{\vert}{\beta}_{i}{\vert}},
  1/b_{2},1/b_{3})
   \label{tef}, \\
   t_{i=g,h}&=&
  {\max}(
   \sqrt{{\alpha}_{a}},
   \sqrt{{\vert}{\beta}_{i}{\vert}},
  1/b_{1},1/b_{2})
   \label{tgh},
   \end{eqnarray}
 where ${\alpha}_{e}$ and ${\alpha}_{a}$ are the virtuality
 of the internal gluon of emission and annihilation diagrams,
 respectively.
 The subscript on ${\beta}_{i}$, the virtuality of the internal
 quark, corresponds to one index of Fig.\ref{fig01}.
 Their expressions are
   \begin{eqnarray}
  {\alpha}_{e}&=&
   \bar{x}_{1}\bar{x}_{2}t
  -\bar{x}_{1}^{2}m_{1}^{2}
  -\bar{x}_{2}^{2}m_{2}^{2}
   >0
   \label{alphae}, \\
  {\alpha}_{a}&=&
   x_{2}\bar{x}_{3}s
  +x_{2}^{2}m_{2}^{2}
  +\bar{x}_{3}^{2}m_{3}^{2}
   >0
   \label{alphaa}, \\
  {\beta}_{a}&=&
   \bar{x}_{2}t
  -\bar{x}_{2}^{2}m_{2}^{2}
  -m_{1}^{2}+m_{c}^{2}
   >0
   \label{betaa}, \\
  {\beta}_{b}&=&
   \bar{x}_{1}t
  -\bar{x}_{1}^{2}m_{1}^{2}
  -m_{2}^{2}
   >0
   \label{betab}, \\
  {\beta}_{c}&=&
   x_{1}x_{2}t
  +x_{1}x_{3}u
  -x_{2}x_{3}s
   \nonumber \\ &-&
   x_{1}^{2}m_{1}^{2}
  -x_{2}^{2}m_{2}^{2}
  -x_{3}^{2}m_{3}^{2}
   \label{betac}, \\
  {\beta}_{d}&=&
   x_{1}x_{2}t
  +x_{1}\bar{x}_{3}u
  -x_{2}\bar{x}_{3}s
   \nonumber \\ &-&
   x_{1}^{2}m_{1}^{2}
  -x_{2}^{2}m_{2}^{2}
  -\bar{x}_{3}^{2}m_{3}^{2}
   \label{betad}, \\
  {\beta}_{e}&=&
   m_{3}^{2}
  +x_{2}^{2}m_{2}^{2}
  +x_{2}s
   >0
   \label{betae}, \\
  {\beta}_{f}&=&
   m_{b}^{2}
  +x_{3}u-m_{1}^{2}
  -x_{3}^{2}m_{3}^{2}
   \label{betaf}, \\
  {\beta}_{g}&=&
   x_{1}\bar{x}_{2}t
  +x_{1}x_{3}u
  -\bar{x}_{2}x_{3}s
   \nonumber \\ &+&
   m_{b}^{2}
  -x_{1}^{2}m_{1}^{2}
  -\bar{x}_{2}^{2}m_{2}^{2}
  -x_{3}^{2}m_{3}^{2}
   \label{betag}, \\
  {\beta}_{h}&=&
   x_{1}x_{2}t
  +x_{1}\bar{x}_{3}u
  -x_{2}\bar{x}_{3}s
  \nonumber \\ &+&
   m_{c}^{2}
  -x_{1}^{2}m_{1}^{2}
  -x_{2}^{2}m_{2}^{2}
  -\bar{x}_{3}^{2}m_{3}^{2}
   \label{betah}.
   \end{eqnarray}
  \section{decay amplitudes}
  \label{app04}
   \begin{equation}
  {\cal A}(B_{c}^{+}{\to}B_{s}^{0}{\pi}^{+})
   = V_{ud}V_{cs}^{\ast} \{
      a_{1}M^{P}_{ab,1}
     +C_{2}M^{P}_{cd,1} \}
   \label{bspi},
   \end{equation}
   \begin{equation}
  {\cal A}(B_{c}^{+}{\to}B_{s}^{0}{\rho}^{+})
   = V_{ud}V_{cs}^{\ast} \{
      a_{1}M^{V}_{ab,1}
     +C_{2}M^{V}_{cd,1} \}
   \label{bsrho},
   \end{equation}
   \begin{eqnarray}
  {\cal A}(B_{c}^{+}{\to}B_{s}^{0}K^{+})
   &=&V_{us}V_{cs}^{\ast}\{
      a_{1}M^{P}_{ab,1}
     +C_{2}M^{P}_{cd,1} \}
    - V_{ub}V_{cb}^{\ast}\{
      (a_{4}-a_{10}/2)M^{P}_{ab,1}
   \nonumber \\ &+&
      (a_{6}-a_{8}/2)M^{P}_{ab,3}
     +(C_{3}-C_{9}/2)M^{P}_{cd,1}
     +(C_{5}-C_{7}/2)M^{P}_{cd,3}
   \nonumber \\ &-&
      a_{1}M^{P}_{ef,1}
     -C_{2}M^{P}_{gh,1} \}
   \label{bskp},
   \end{eqnarray}
   \begin{eqnarray}
  {\cal A}(B_{c}^{+}{\to}B_{s}^{0}K^{{\ast}+})
   &=&V_{us}V_{cs}^{\ast}\{
      a_{1}M^{V}_{ab,1}
     +C_{2}M^{V}_{cd,1} \}
   \nonumber \\ &-&
      V_{ub}V_{cb}^{\ast}\{
      (a_{4}-a_{10}/2)M^{V}_{ab,1}
     +(C_{3}-C_{9}/2)M^{V}_{cd,1}
   \nonumber \\ & &
     +(C_{5}-C_{7}/2)M^{V}_{cd,3}
     -a_{1}M^{V}_{ef,1}
     -C_{2}M^{V}_{gh,1} \}
   \label{bskv},
   \end{eqnarray}
   \begin{eqnarray}
  {\cal A}(B_{c}^{+}{\to}B_{d}^{0}{\pi}^{+})
   &=&V_{ud}V_{cd}^{\ast}\{
      a_{1}M^{P}_{ab,1}
     +C_{2}M^{P}_{cd,1} \}
    - V_{ub}V_{cb}^{\ast}\{
      (a_{4}-a_{10}/2)M^{P}_{ab,1}
   \nonumber \\ &+&
      (a_{6}-a_{8}/2)M^{P}_{ab,3}
     +(C_{3}-C_{9}/2)M^{P}_{cd,1}
     +(C_{5}-C_{7}/2)M^{P}_{cd,3}
   \nonumber \\ &-&
      a_{1}M^{P}_{ef,1}
     -C_{2}M^{P}_{gh,1} \}
   \label{bdpi},
   \end{eqnarray}
   \begin{eqnarray}
  {\cal A}(B_{c}^{+}{\to}B_{d}^{0}{\rho}^{+})
   &=&V_{ud}V_{cd}^{\ast}\{
      a_{1}M^{V}_{ab,1}
     +C_{2}M^{V}_{cd,1} \}
   \nonumber \\ &-&
      V_{ub}V_{cb}^{\ast}\{
      (a_{4}-a_{10}/2)M^{V}_{ab,1}
     +(C_{3}-C_{9}/2)M^{V}_{cd,1}
   \nonumber \\ & &
     +(C_{5}-C_{7}/2)M^{V}_{cd,3}
     -a_{1}M^{V}_{ef,1}
     -C_{2}M^{V}_{gh,1} \}
   \label{bdrho},
   \end{eqnarray}
   \begin{equation}
  {\cal A}(B_{c}^{+}{\to}B_{d}^{0}K^{+})
   =  V_{us}V_{cd}^{\ast}\{
      a_{1}M^{P}_{ab,1}
     +C_{2}M^{P}_{cd,1} \}
   \label{bdkp},
   \end{equation}
   \begin{equation}
  {\cal A}(B_{c}^{+}{\to}B_{d}^{0}K^{{\ast}+})
   =  V_{us}V_{cd}^{\ast}\{
      a_{1}M^{V}_{ab,1}
     +C_{2}M^{V}_{cd,1} \}
   \label{bdkv},
   \end{equation}
   \begin{equation}
  {\cal A}(B_{c}^{+}{\to}B_{u}^{+}\overline{K}^{0})
   =  V_{ud}V_{cs}^{\ast}\{
      a_{2}M^{P}_{ab,1}
     +C_{1}M^{P}_{cd,1} \}
   \label{bukb},
   \end{equation}
   \begin{equation}
  {\cal A}(B_{c}^{+}{\to}B_{u}^{+}\overline{K}^{{\ast}0})
   =  V_{ud}V_{cs}^{\ast}\{
      a_{2}M^{V}_{ab,1}
     +C_{1}M^{V}_{cd,1} \}
   \label{bukvb},
   \end{equation}
   \begin{equation}
  {\cal A}(B_{c}^{+}{\to}B_{u}^{+}K^{0})
   =  V_{us}V_{cd}^{\ast}\{
      a_{2}M^{P}_{ab,1}
     +C_{1}M^{P}_{cd,1} \}
   \label{bukz},
   \end{equation}
   \begin{equation}
  {\cal A}(B_{c}^{+}{\to}B_{u}^{+}K^{{\ast}0})
   =  V_{us}V_{cd}^{\ast}\{
      a_{2}M^{V}_{ab,1}
     +C_{1}M^{V}_{cd,1} \}
   \label{bukvz},
   \end{equation}
   \begin{eqnarray}
    \sqrt{2}
   {\cal A}(B_{c}^{+}{\to}B_{u}^{+}{\pi}^{0})
  &=&-V_{ud}V_{cd}^{\ast}\{
      a_{2}M^{P}_{ab,1}
     +C_{1}M^{P}_{cd,1} \}
   -  V_{ub}V_{cb}^{\ast}\{
     -a_{1}M^{P}_{ef,1}
     -C_{2}M^{P}_{gh,1}
   \nonumber \\ &+&
      (a_{4}+a_{10}+\frac{3}{2}a_{9})M^{P}_{ab,1}
     +\frac{3}{2}a_{7}M^{P}_{ab,2}
     +(a_{6}+a_{8})M^{P}_{ab,3}
   \nonumber \\ &+&
      (C_{3}+C_{9}+\frac{3}{2}C_{10})M^{P}_{cd,1}
     +\frac{3}{2}C_{8}M^{P}_{cd,2}
     +(C_{5}+C_{7})M^{P}_{cd,3} \}
   \label{bupi},
   \end{eqnarray}
   \begin{eqnarray}
    \sqrt{2}
   {\cal A}(B_{c}^{+}{\to}B_{u}^{+}{\rho}^{0})
  &=&-V_{ud}V_{cd}^{\ast}\{
      a_{2}M^{V}_{ab,1}
     +C_{1}M^{V}_{cd,1} \}
     -V_{ub}V_{cb}^{\ast}\{
      (C_{5}+C_{7})M^{V}_{cd,3}
   \nonumber \\ & &
     -a_{1}M^{V}_{ef,1}
     +(a_{4}+a_{10}+\frac{3}{2}a_{9})M^{V}_{ab,1}
     +\frac{3}{2}a_{7}M^{V}_{ab,2}
   \nonumber \\ & &
     -C_{2}M^{V}_{gh,1}
     +(C_{3}+C_{9}+\frac{3}{2}C_{10})M^{V}_{cd,1}
     +\frac{3}{2}C_{8}M^{V}_{cd,2} \}
   \label{burho},
   \end{eqnarray}
   \begin{eqnarray}
    \sqrt{2}
   {\cal A}(B_{c}^{+}{\to}B_{u}^{+}{\omega})
  &=& V_{ud}V_{cd}^{\ast}\{
      a_{2}M^{V}_{ab,1}
     +C_{1}M^{V}_{cd,1} \}
     -V_{ub}V_{cb}^{\ast}\{
     (C_{5}+C_{7})M^{V}_{cd,3}
   \nonumber \\ & &
    +(2a_{3}+a_{4}+a_{9}/2+a_{10})M^{V}_{ab,1}
    +(2a_{5}+a_{7}/2)M^{V}_{ab,2}
   \nonumber \\ & &
    +(C_{3}+2C_{4}+C_{9}+C_{10}/2)M^{V}_{cd,1}
    +(2C_{6}+C_{8}/2)M^{V}_{cd,2}
   \nonumber \\ & &
    -a_{1}M^{V}_{ef,1}
    -C_{2}M^{V}_{gh,1} \}
   \label{buomega},
   \end{eqnarray}
   \begin{eqnarray}
    \sqrt{2}
   {\cal A}(B_{c}^{+}{\to}B_{u}^{+}{\eta}_{q})
  &=& V_{ud}V_{cd}^{\ast}\{
      a_{2}M^{P}_{ab,1}
     +C_{1}M^{P}_{cd,1} \}
   -  V_{ub}V_{cb}^{\ast}\{
   -  a_{1}M^{P}_{ef,1}
   -  C_{2}M^{P}_{gh,1}
   \nonumber \\ & &
    +(2a_{3}+a_{4}+a_{9}/2+a_{10})M^{P}_{ab,1}
    +(2a_{5}+a_{7}/2)M^{P}_{ab,2}
   \nonumber \\ & &
    +(C_{3}+2C_{4}+C_{9}+C_{10}/2)M^{P}_{cd,1}
    +(2C_{6}+C_{8}/2)M^{P}_{cd,2}
   \nonumber \\ & &
    +(a_{6}+a_{8})M^{P}_{ab,3}
    +(C_{5}+C_{7})M^{P}_{cd,3} \}
   \label{buetaq}.
   \end{eqnarray}
   \begin{eqnarray}
   {\cal A}(B_{c}^{+}{\to}B_{u}^{+}{\eta}_{s})
  &=& V_{us}V_{cs}^{\ast}\{
      a_{2}M^{P}_{ab,1}
     +C_{1}M^{P}_{cd,1} \}
   \nonumber \\ &-&
       V_{ub}V_{cb}^{\ast}\{
      (a_{3}-\frac{1}{2}a_{9})M^{P}_{ab,1}
     +(a_{5}-\frac{1}{2}a_{7})M^{P}_{ab,2}
   \nonumber \\ & &
     +(C_{4}-\frac{1}{2}C_{10})M^{P}_{cd,1}
     +(C_{6}-\frac{1}{2}C_{8})M^{P}_{cd,2} \}
   \label{buetas},
   \end{eqnarray}
   \begin{eqnarray}
  {\cal A}(B_{c}^{+}{\to}B_{u}^{+}{\eta})
  &=&
  {\cos}{\phi}{\cal A}(B_{c}^{+}{\to}B_{u}^{+}{\eta}_{q})
 -{\sin}{\phi}{\cal A}(B_{c}^{+}{\to}B_{u}^{+}{\eta}_{s})
   \label{bueta}, \\
  {\cal A}(B_{c}^{+}{\to}B_{u}^{+}{\eta}^{\prime})
  &=&
  {\sin}{\phi}{\cal A}(B_{c}^{+}{\to}B_{u}^{+}{\eta}_{q})
 +{\cos}{\phi}{\cal A}(B_{c}^{+}{\to}B_{u}^{+}{\eta}_{s})
   \label{buetap},
   \end{eqnarray}

  \end{appendix}

  {{\renewcommand{\baselinestretch}{0.9}
  \begin{sidewaystable}[h]
 {\scriptsize
  \caption{Branching ratios of $B_{c}$ ${\to}$ $BP$, $BV$
  decays with the fixed coefficients $a_{1}$ = 1.20 and
  $a_{2}$ = $-0.317$, and form factors
  $F_{0}^{B_{c}{\to}B_{q}}(0)$.}
  \label{tab01}
  \begin{tabular}{l|r|r|r|r|r|r|r|r|r|r|r|r} \hline \hline
     reference
   & \cite{prd.86.094028}\footnotemark[1]
   & \cite{prd.80.114003}\footnotemark[2]
   & \cite{prd.77.114004}\footnotemark[3]
   & \cite{prd.74.074008}\footnotemark[4]
   & \cite{prd.73.054024}\footnotemark[5]
   & \cite{epjc.32.29}\footnotemark[6]
   & \cite{pan.67.1559}\footnotemark[7]
   & \cite{prd.61.034012}\footnotemark[8]
   & \cite{prd.39.1342}\footnotemark[9]
   & \cite{prd.49.3399}\footnotemark[10]
   & \cite{prd.62.014019}\footnotemark[11]
   & \cite{zpc.51.549}\footnotemark[12] \\ \hline
     $F_{0}^{B_{c}{\to}B}(0)$
   & 1.01
   & 0.467 (0.426)
   & 0.8
   & 0.39
   & 0.58
   & 0.39
   & 1.27
   & 0.66
   & 0.831
   & & &
   \\ \hline
     $F_{0}^{B_{c}{\to}B_{s}}(0)$
   & 1.03
   & 0.573 (0.571)
   & 0.8
   & 0.58
   & 0.61
   & 0.50
   & 1.3
   & 0.66
   & 0.859
   & & &
   \\ \hline
     ${\cal B}r(B_{c}^{+}{\to}B_{s}^{0}{\pi}^{+})$
   & 10.9${\times}10^{-2}$
   & 3.72 (3.70)${\times}10^{-2}$
   & 5.31${\times}10^{-2}$
   & 3.51${\times}10^{-2}$
   & 3.9${\times}10^{-2}$
   & 2.52${\times}10^{-2}$
   & 16.4${\times}10^{-2}$
   & 3.03${\times}10^{-2}$
   & 7.85${\times}10^{-2}$
   & 5.79${\times}10^{-2}$
   & 1.57${\times}10^{-2}$
   & 3.08 (4.36)${\times}10^{-2}$
   \\ \hline
     ${\cal B}r(B_{c}^{+}{\to}B_{s}^{0}{\rho}^{+})$
   & 9.05${\times}10^{-2}$
   & 2.56 (2.34)${\times}10^{-2}$
   & 6.27${\times}10^{-2}$
   & 2.34${\times}10^{-2}$
   & 2.3${\times}10^{-2}$
   & 1.41${\times}10^{-2}$
   & 7.2${\times}10^{-2}$
   & 1.35${\times}10^{-2}$
   & 4.70${\times}10^{-2}$
   & 4.44${\times}10^{-2}$
   & 3.88${\times}10^{-2}$
   & 1.24 (2.00)${\times}10^{-2}$
   \\ \hline
     ${\cal B}r(B_{c}^{+}{\to}B_{s}^{0}K^{+})$
   & 7.23${\times}10^{-3}$
   & 2.87 (2.84)${\times}10^{-3}$
   & 3.68${\times}10^{-3}$
   & 2.9${\times}10^{-3}$
   & 2.9${\times}10^{-3}$
   & 2.1${\times}10^{-3}$
   & 1.06${\times}10^{-2}$
   & 2.13${\times}10^{-3}$
   & 5.71${\times}10^{-3}$
   & 4.16${\times}10^{-3}$
   & 1.68${\times}10^{-3}$
   & 2.16 (3.25)${\times}10^{-3}$
   \\ \hline
     ${\cal B}r(B_{c}^{+}{\to}B_{s}^{0}K^{{\ast}+})$
   & 3.4${\times}10^{-4}$
   & 6.9 (6.1)${\times}10^{-5}$
   & 1.65${\times}10^{-3}$
   & 1.3${\times}10^{-4}$
   & 1.1${\times}10^{-4}$
   & 3.0${\times}10^{-5}$
   &
   & 4.26${\times}10^{-5}$
   & 2.36${\times}10^{-4}$
   & 2.93${\times}10^{-3}$
   & 1.05${\times}10^{-3}$
   &
   \\ \hline
     ${\cal B}r(B_{c}^{+}{\to}B_{d}^{0}{\pi}^{+})$
   & 7.2${\times}10^{-3}$
   & 1.57 (1.31)${\times}10^{-3}$
   & 3.73${\times}10^{-3}$
   & 1.1${\times}10^{-3}$
   & 2.0${\times}10^{-3}$
   & 1.0${\times}10^{-3}$
   & 1.06${\times}10^{-2}$
   & 1.95${\times}10^{-3}$
   & 5.35${\times}10^{-3}$
   & 3.27${\times}10^{-3}$
   & 1.02${\times}10^{-3}$
   & 0.96 (1.87)${\times}10^{-3}$
   \\ \hline
     ${\cal B}r(B_{c}^{+}{\to}B_{d}^{0}{\rho}^{+})$
   & 1.18${\times}10^{-2}$
   & 1.95 (1.52)${\times}10^{-3}$
   & 5.27${\times}10^{-3}$
   & 1.4${\times}10^{-3}$
   & 2.0${\times}10^{-3}$
   & 1.3${\times}10^{-3}$
   & 9.6${\times}10^{-3}$
   & 1.53${\times}10^{-3}$
   & 5.98${\times}10^{-3}$
   & 5.92${\times}10^{-3}$
   & 2.78${\times}10^{-3}$
   & 0.93 (2.12)${\times}10^{-3}$
   \\ \hline
     ${\cal B}r(B_{c}^{+}{\to}B_{d}^{0}K^{+})$
   & 5.4${\times}10^{-4}$
   & 1.3 (1.1)${\times}10^{-4}$
   & 2.66${\times}10^{-4}$
   & 1.0${\times}10^{-4}$
   & 1.5${\times}10^{-4}$
   & 9.0${\times}10^{-5}$
   & 7.0${\times}10^{-4}$
   & 1.39${\times}10^{-4}$
   &
   & 2.53${\times}10^{-4}$
   & 1.04${\times}10^{-4}$
   &
   \\ \hline
     ${\cal B}r(B_{c}^{+}{\to}B_{d}^{0}K^{{\ast}+})$
   & 2.9${\times}10^{-4}$
   & 4.2 (3.2)${\times}10^{-5}$
   & 2.26${\times}10^{-4}$
   & 3.9${\times}10^{-5}$
   & 4.8${\times}10^{-5}$
   & 4.0${\times}10^{-5}$
   & 1.5${\times}10^{-4}$
   & 3.17${\times}10^{-5}$
   &
   & 1.78${\times}10^{-4}$
   & 1.24${\times}10^{-4}$
   &
   \\ \hline
     ${\cal B}r(B_{c}^{+}{\to}B_{u}^{+}\overline{K}^{0})$
   & 1.26${\times}10^{-2}$
   & 3.36 (2.79)${\times}10^{-3}$
   & 2.21${\times}10^{-5}$
   & 2.5${\times}10^{-3}$
   & 3.8${\times}10^{-3}$
   & 2.4${\times}10^{-3}$
   & 1.98${\times}10^{-2}$
   &
   & 1.72${\times}10^{-2}$
   & 6.67${\times}10^{-3}$
   & 2.70${\times}10^{-3}$
   & 1.95 (4.25)${\times}10^{-3}$
   \\ \hline
     ${\cal B}r(B_{c}^{+}{\to}B_{u}^{+}\overline{K}^{{\ast}0})$
   & 7.1${\times}10^{-3}$
   & 1.08 (0.80)${\times}10^{-3}$
   & 1.84${\times}10^{-5}$
   & 9.3${\times}10^{-4}$
   & 1.1${\times}10^{-3}$
   & 9.0${\times}10^{-4}$
   & 4.3${\times}10^{-3}$
   &
   & 6.30${\times}10^{-3}$
   & 4.72${\times}10^{-3}$
   & 3.24${\times}10^{-3}$
   & 0.69 (1.67)${\times}10^{-3}$
   \\ \hline
     ${\cal B}r(B_{c}^{+}{\to}B_{u}^{+}{\pi}^{0})$
   & 2.5${\times}10^{-4}$
   & 5.5 (4.6)${\times}10^{-5}$
   & 4.51${\times}10^{-7}$
   & 3.8${\times}10^{-5}$
   & 7.0${\times}10^{-5}$
   & 4.0${\times}10^{-5}$
   & 3.7${\times}10^{-4}$
   &
   & 3.23${\times}10^{-4}$
   & 1.14${\times}10^{-4}$
   & 3.53${\times}10^{-5}$
   & 3.32 (6.57)${\times}10^{-5}$
   \\ \hline
     ${\cal B}r(B_{c}^{+}{\to}B_{u}^{+}{\rho}^{0})$
   & 4.1${\times}10^{-4}$
   & 6.8 (5.3)${\times}10^{-5}$
   & 6.48${\times}10^{-7}$
   & 5.0${\times}10^{-5}$
   & 7.1${\times}10^{-5}$
   & 5.0${\times}10^{-5}$
   & 3.4${\times}10^{-4}$
   &
   & 3.59${\times}10^{-4}$
   & 2.06${\times}10^{-4}$
   & 9.68${\times}10^{-5}$
   & 3.25 (7.40)${\times}10^{-5}$
    \\ \hline
     ${\cal B}r(B_{c}^{+}{\to}B_{u}^{+}{\omega})$
   &
   & 5.1 (3.9)${\times}10^{-5}$
   & 5.82${\times}10^{-7}$
   & & & & &
   & 3.36${\times}10^{-4}$
   & &
   & 2.63 (6.02)${\times}10^{-5}$
   \\ \hline
     ${\cal B}r(B_{c}^{+}{\to}B_{u}^{+}{\eta})$
   &
   & 2.8 (2.3)${\times}10^{-4}$
   & 1.61${\times}10^{-6}$
   & & & & & & & & \\ \hline
     ${\cal B}r(B_{c}^{+}{\to}B_{u}^{+}{\eta}^{\prime})$
   &
   & 3.8 (3.2)${\times}10^{-6}$
   & 8.77${\times}10^{-8}$
   & & & & & & & & \\ \hline
     ${\cal B}r(B_{c}^{+}{\to}B_{u}^{+}K^{0})$
   &
   & 8.8 (7.3)${\times}10^{-6}$
   & 6.54${\times}10^{-8}$
   & & & & & & &
   & \\ \hline
     ${\cal B}r(B_{c}^{+}{\to}B_{u}^{+}K^{{\ast}0})$
   &
   & 2.8 (2.1)${\times}10^{-6}$
   & 5.47${\times}10^{-8}$
   & & & & & & &
   & \\ \hline \hline
   \end{tabular} }
   \footnotetext[1]{It is estimated in the relativistic
    independent quark model based on the scalar-vector
    form confining potential.}
   \footnotetext[2]{It is estimated in the light-front
    quark model using the Coulomb plus linear confining
    (harmonic oscillator) potential.}
   \footnotetext[3]{It is estimated at the leading order in
    the QCD factorziation approach with Wilson coefficients
    $c_{1}$ $=$ 1.22 and $c_{2}$ $=$ $-0.42$.}
   \footnotetext[4]{It is estimated in the nonrelativistic
    constituent quark model using the Coulomb plus confining
    potential.}
   \footnotetext[5]{It is estimated in the relativistic
    constituent quark model.}
   \footnotetext[6]{It is estimated in the relativistic
    constituent quark model.}
   \footnotetext[7]{It is estimated in the QCD sum rules.}
   \footnotetext[8]{It is estimated in the constituent
    quark model.}
   \footnotetext[9]{It is estimated in the BSW model with
    ${\omega}$ = 0.8 GeV.}
   \footnotetext[10]{It is estimated in the potential model
    based on the Bethe-Salpeter equation.}
   \footnotetext[11]{It is estimated in the relativistic model
    based on the Bethe-Salpeter equation.}
   \footnotetext[12]{It is estimated in the BSW (ISGW) model.}
   \end{sidewaystable}}

  \begin{table}[h]
  \caption{Numerical values of Wilson coefficients
   at different scales.}
  \label{tab02}
  \begin{tabular}{l|rrrr} \hline
  ${\mu}$ & 1 GeV & $m_{c}$ & 2 GeV & $m_{b}$ \\ \hline
  $C_{1}$ & $ 1.294 $ & $ 1.230 $ & $ 1.156 $ & $ 1.087 $ \\
  $C_{2}{\times}10$ & $-5.327$ & $-4.370$ & $-3.177$ & $-1.947$ \\
  $C_{3}{\times}10^{2}$ & $ 4.764$ & $ 3.639$ & $ 2.471$ & $ 1.482$ \\
  $C_{4}{\times}10^{2}$ & $-9.674$ & $-7.731$ & $-5.602$ & $-3.605$ \\
  $C_{5}{\times}10^{3}$ & $ 7.009$ & $ 9.963$ & $ 10.55$ & $ 8.613$ \\
  $C_{6}{\times}10^{2}$ & $-15.50$ & $-11.31$ & $-7.339$ & $-4.240$ \\
  $C_{7}{\times}10^{5}$ & $-7.465$ & $-11.53$ & $-10.98$ & $0.4438$ \\
  $C_{8}{\times}10^{3}$ & $ 1.660$ & $ 1.205$ & $0.7759$ & $0.4491$ \\
  $C_{9}{\times}10^{2}$ & $-1.213$ & $-1.149$ & $-1.078$ & $-1.009$ \\
  $C_{10}{\times}10^{3}$ & $5.493$ & $ 4.474$ & $ 3.287$ & $ 2.131$ \\ \hline
  \end{tabular}
  \end{table}

  \begin{table}[h]
  \caption{Numerical values of the input parameters.}
  \label{tab03}
  \begin{ruledtabular}
  \begin{tabular}{ll}
  \multicolumn{2}{c}{Wolfenstein parameters} \\ \hline
    ${\lambda}$ $=$ $0.22535{\pm}0.00065$ \cite{pdg.2012}
  & $A$ $=$ $0.811^{+0.022}_{-0.012}$ \cite{pdg.2012} \\
    $\bar{\rho}$ $=$ $0.131^{+0.026}_{-0.013}$ \cite{pdg.2012}
  & $\bar{\eta}$ $=$ $0.345^{+0.013}_{-0.014}$ \cite{pdg.2012} \\ \hline
  \multicolumn{2}{c}{masses of mesons and quarks} \\ \hline
    $m_{B_{u}}$ $=$ $5279.25{\pm}0.17$~MeV \cite{pdg.2012}
  & $m_{B_{d}}$ $=$ $5279.58{\pm}0.17$~MeV \cite{pdg.2012} \\
    $m_{B_{s}}$ $=$ $5366.77{\pm}0.24$~MeV \cite{pdg.2012}
  & $m_{B_{c}}$ $=$ $6.277{\pm}0.006$~GeV \cite{pdg.2012} \\
    $m_{c}$ $=$ $1.275{\pm}0.025$ GeV \cite{pdg.2012}
  & $m_{b}$ $=$ $4.18 {\pm}0.03 $ GeV \cite{pdg.2012} \\ \hline
  \multicolumn{2}{c}{decay constant of mesons} \\ \hline
    $f_{\pi}$      $=$ $130.41{\pm}0.20$ MeV \cite{pdg.2012}
  & $f_{K}  $      $=$ $156.1{\pm}0.8$ MeV \cite{pdg.2012} \\
    $f_{q}$ $=$ $(1.07{\pm}0.02)f_{\pi}$ \cite{prd.58.114006}
  & $f_{s}$ $=$ $(1.34{\pm}0.06)f_{\pi}$ \cite{prd.58.114006} \\
    $f_{B_{u,d}}$ $=$ $190.5{\pm}4.2$ MeV \cite{1310.8555}
  & $f_{B_{s}}$   $=$ $227.7{\pm}4.5$ MeV \cite{1310.8555} \\
    $f_{\rho}$     $=$ $216{\pm}3$ MeV \cite{JHEP.03.069.2007}
  & $f_{\rho}^{T}(1\,{\rm GeV})$   $=$ $165{\pm}9$ MeV \cite{JHEP.03.069.2007} \\
    $f_{\omega}$   $=$ $187{\pm}5$ MeV \cite{JHEP.03.069.2007}
  & $f_{\omega}^{T}(1\,{\rm GeV})$ $=$ $151{\pm}9$ MeV \cite{JHEP.03.069.2007} \\
    $f_{K^{\ast}}$ $=$ $220{\pm}5$ MeV \cite{JHEP.03.069.2007}
  & $f_{K^{\ast}}^{T}(1\,{\rm GeV})$ $=$ $185{\pm}10$ MeV \cite{JHEP.03.069.2007} \\
    $f_{B_{c}}$ $=$ $489{\pm}4{\pm}3$ MeV \cite{plb.651.171}
  & $f_{3P}(1\,{\rm GeV})$ $=$ $(4.5{\pm}1.5){\times}10^{-3}$ ${\rm GeV}^{2}$ \cite{JHEP.05.004.2006} \\ \hline
  \multicolumn{2}{c}{Gegenbauer moments\footnotemark[1] at the scale ${\mu}$ $=$ 1 GeV}
  \footnotetext[1]{We will take the approximation
    $a_{i}^{{\eta}_{q}}$ $=$ $a_{i}^{{\eta}_{s}}$ $=$ $a_{i}^{\pi}$,
    and $a_{i,{\omega}}^{\parallel}$ $=$ $a_{i,{\rho}}^{\parallel}$} \\ \hline
    $a_{1,{\rho}}^{\parallel}$ $=$ $0$ \cite{JHEP.03.069.2007}
  & $a_{2,{\rho}}^{\parallel}$ $=$ $0.15{\pm}0.07$ \cite{JHEP.03.069.2007} \\
    $a_{1,K^{\ast}}^{\parallel}$ $=$ $0.03{\pm}0.02$ \cite{JHEP.03.069.2007}
  & $a_{2,K^{\ast}}^{\parallel}$ $=$ $0.11{\pm}0.09$ \cite{JHEP.03.069.2007} \\
    $a_{1}^{\pi}$ $=$ $0$ \cite{JHEP.05.004.2006}
  & $a_{2}^{\pi}$ $=$ $0.25{\pm}0.15$ \cite{JHEP.05.004.2006} \\
    $a_{1}^{K}$   $=$ $0.06{\pm}0.03$ \cite{JHEP.05.004.2006}
  & $a_{2}^{K}$   $=$ $0.25{\pm}0.15$ \cite{JHEP.05.004.2006} \\
    ${\omega}_{3}^{\pi}$ $=$ $-1.5{\pm}0.7$ \cite{JHEP.05.004.2006}
  & ${\omega}_{3}^{K}$ $=$ $-1.2{\pm}0.7$ \cite{JHEP.05.004.2006}
  \end{tabular}
  \end{ruledtabular}
  \end{table}

  \begin{table}[h]
  \label{tab04}
  \caption{Form factor and the fitted parameters,
    where the uncertainties are from mass $m_{b}$, $m_{c}$,
    shape parameters ${\omega}_{B_{c}}$, ${\omega}_{B_{q}}$
    and typical scale $t$, respectively.}
    \begin{tabular}{c|c|c|c|c} \hline
    \multirow{3}{*}{$B_{c}$ ${\to}$ $B_{u}$}
  & $F_{0}(0)$
  & $1.074^{+0.007+0.016+0.031+0.172+0.131}_{-0.006-0.017-0.028-0.150-0.056}$
  & $F_{1}(0)$
  & $1.074^{+0.007+0.016+0.031+0.172+0.131}_{-0.006-0.017-0.028-0.150-0.056}$
    \\ \cline{2-5}
  & $m$
  & $1.123^{+0.003+0.001+0.010+0.040+0.021}_{-0.002-0.001-0.010-0.037-0.013}$
  & $m$
  & $1.110^{+0.004+0.011+0.014+0.007+0.022}_{-0.002-0.009-0.014-0.005-0.008}$
    \\ \cline{2-5}
  & ${\delta}$
  & $2.689^{+0.040+0.212+0.104+0.858+0.358}_{-0.027-0.185-0.103-0.658-0.743}$
  & ${\delta}$
  & $1.830^{+0.029+0.092+0.082+0.350+0.251}_{-0.022-0.084-0.083-0.309-0.564}$
    \\ \hline
    \multirow{3}{*}{$B_{c}$ ${\to}$ $B_{d}$}
  & $F_{0}(0)$
  & $1.075^{+0.006+0.016+0.031+0.172+0.131}_{-0.007-0.017-0.028-0.150-0.056}$
  & $F_{1}(0)$
  & $1.075^{+0.006+0.016+0.031+0.172+0.131}_{-0.007-0.017-0.028-0.150-0.056}$
    \\ \cline{2-5}
  & $m$
  & $1.123^{+0.002+0.000+0.009+0.039+0.022}_{-0.002-0.000-0.011-0.038-0.014}$
  & $m$
  & $1.109^{+0.003+0.011+0.013+0.007+0.022}_{-0.003-0.009-0.015-0.006-0.009}$
    \\ \cline{2-5}
  & ${\delta}$
  & $2.691^{+0.032+0.205+0.099+0.849+0.360}_{-0.032-0.191-0.111-0.664-0.749}$
  & ${\delta}$
  & $1.831^{+0.025+0.088+0.079+0.346+0.251}_{-0.024-0.087-0.086-0.312-0.566}$
    \\ \hline
    \multirow{3}{*}{$B_{c}$ ${\to}$ $B_{s}$}
  & $F_{0}(0)$
  & $1.034^{+0.008+0.014+0.035+0.177+0.141}_{-0.008-0.015-0.031-0.154-0.058}$
  & $F_{1}(0)$
  & $1.034^{+0.008+0.014+0.035+0.177+0.141}_{-0.008-0.015-0.031-0.154-0.058}$
    \\ \cline{2-5}
  & $m$
  & $1.224^{+0.004+0.019+0.009+0.101+0.044}_{-0.004-0.018-0.010-0.081-0.058}$
  & $m$
  & $1.065^{+0.003+0.007+0.010+0.038+0.028}_{-0.003-0.005-0.011-0.032-0.030}$
    \\ \cline{2-5}
  & ${\delta}$
  & $6.005^{+0.092+0.161+0.179+3.239+1.193}_{-0.091-0.149-0.190-1.963-2.141}$
  & ${\delta}$
  & $3.176^{+0.045+0.050+0.099+0.887+0.482}_{-0.044-0.044-0.107-0.673-0.982}$
    \\ \hline
  \end{tabular}
  \end{table}

  \begin{sidewaystable}[h]
  \caption{branching ratio for the $B_{c}$ ${\to}$ $BP$, $BV$
    decays, where ${\cal B}^{\rm t}$ denote the contributions
    from only the tree operators, ${\cal B}^{\rm t+p}$ denote
    the contributions from both the tree and penguin operators,
    and ${\cal B}^{\rm t+p+a}$ denote
    the contributions of the tree, penguin, and annihilation
    topologies; the uncertainties are from mass $m_{b}$,
    $m_{c}$, shape parameters ${\omega}_{B_{c}}$,
    ${\omega}_{B_{q}}$ and typical scale $t$, respectively.}
    \label{tab05}
    \begin{tabular}{c|c|c|c} \hline
     mode
   & ${\cal B}^{\rm t}$
   & ${\cal B}^{\rm t+p}$
   & ${\cal B}^{\rm t+p+a}$ \\ \hline
     $B_{s}^{0}{\pi}^{+}$
   & $8.822^{+0.145+0.120+0.631+3.448+3.178}_{-0.074-0.024-0.526-3.658-1.334}{\times}10^{-2}$
   & &
    \\ \hline
     $B_{s}^{0}{\rho}^{+}$
   & $3.190^{+0.043+0.041+0.205+1.263+1.123}_{-0.048-0.057-0.192-0.926-0.460}{\times}10^{-2}$
   & & \\ \hline
     $B_{s}^{0}K^{+}$
   & $5.237^{+0.037+0.056+0.308+2.133+1.956}_{-0.112-0.056-0.376-1.591-0.794}{\times}10^{-3}$
   & $5.250^{+0.037+0.056+0.310+2.141+1.968}_{-0.111-0.056-0.377-1.595-0.797}{\times}10^{-3}$
   & $5.441^{+0.037+0.057+0.315+2.239+2.019}_{-0.114-0.057-0.384-1.662-0.821}{\times}10^{-3}$
   \\ \hline
     $B_{s}^{0}K^{{\ast}+}$
   & $9.665^{+0.202+0.200+0.675+3.715+4.775}_{-0.157-0.026-0.594-2.781-1.138}{\times}10^{-5}$
   & $9.671^{+0.199+0.199+0.669+3.719+4.769}_{-0.159-0.031-0.595-2.785-1.142}{\times}10^{-5}$
   & $9.726^{+0.200+0.200+0.674+3.744+4.794}_{-0.159-0.031-0.596-2.803-1.147}{\times}10^{-5}$
   \\ \hline
     $B_{d}^{0}{\pi}^{+}$
   & $6.850^{+0.080+0.208+0.400+2.511+2.242}_{-0.086-0.207-0.329-1.896-0.901}{\times}10^{-3}$
   & $6.833^{+0.081+0.208+0.401+2.505+2.230}_{-0.085-0.205-0.327-1.890-0.896}{\times}10^{-3}$
   & $6.772^{+0.080+0.207+0.398+2.475+2.215}_{-0.085-0.205-0.326-1.870-0.890}{\times}10^{-3}$
   \\ \hline
     $B_{d}^{0}{\rho}^{+}$
   & $4.280^{+0.049+0.053+0.251+1.589+1.418}_{-0.054-0.146-0.214-1.186-0.573}{\times}10^{-3}$
   & $4.279^{+0.049+0.053+0.251+1.589+1.417}_{-0.054-0.146-0.214-1.186-0.573}{\times}10^{-3}$
   & $4.253^{+0.049+0.053+0.251+1.576+1.412}_{-0.054-0.145-0.213-1.177-0.570}{\times}10^{-3}$
   \\ \hline
     $B_{d}^{0}K^{+}$
   & $4.370^{+0.051+0.153+0.254+1.496+1.508}_{-0.054-0.152-0.206-1.269-0.595}{\times}10^{-4}$
   & & \\ \hline
     $B_{d}^{0}K^{{\ast}+}$
   & $8.305^{+0.102+0.301+0.494+3.065+2.775}_{-0.100-0.292-0.440-2.266-1.064}{\times}10^{-5}$
   & & \\ \hline
     $B_{u}^{+}\overline{K}^{0}$
   & $2.205^{+0.012+0.052+0.138+0.773+2.158}_{-0.019-0.100-0.126-0.694-0.993}{\times}10^{-3}$
   & & \\ \hline
     $B_{u}^{+}\overline{K}^{{\ast}0}$
   & $1.958^{+0.021+0.178+0.235+1.308+2.586}_{-0.036-0.038-0.069-0.520-0.893}{\times}10^{-4}$
   & & \\ \hline
     $B_{u}^{+}{\pi}^{0}$
   & $5.222^{+0.131+0.245+0.140+1.963+4.126}_{-0.034-0.311-0.620-1.724-2.361}{\times}10^{-5}$
   & $5.269^{+0.130+0.245+0.141+2.001+4.094}_{-0.033-0.305-0.615-1.740-2.353}{\times}10^{-5}$
   & $4.924^{+0.123+0.235+0.136+1.877+3.928}_{-0.030-0.291-0.589-1.630-2.232}{\times}10^{-5}$
   \\ \hline
     $B_{u}^{+}{\rho}^{0}$
   & $1.838^{+0.036+0.169+0.309+1.218+2.294}_{-0.012-0.070-0.028-0.427-0.812}{\times}10^{-5}$
   & $1.840^{+0.036+0.168+0.308+1.218+2.289}_{-0.012-0.067-0.028-0.427-0.811}{\times}10^{-5}$
   & $1.716^{+0.034+0.162+0.297+1.148+2.211}_{-0.010-0.065-0.027-0.393-0.768}{\times}10^{-5}$
    \\ \hline
     $B_{u}^{+}{\omega}$
   & $1.281^{+0.003+0.165+0.215+0.863+1.673}_{-0.010-0.029-0.023-0.385-0.566}{\times}10^{-5}$
   & $1.280^{+0.004+0.166+0.216+0.864+1.681}_{-0.009-0.030-0.024-0.385-0.567}{\times}10^{-5}$
   & $1.371^{+0.005+0.173+0.225+0.916+1.739}_{-0.010-0.031-0.025-0.413-0.598}{\times}10^{-5}$
   \\ \hline
     $B_{u}^{+}{\eta}$
   & $1.417^{+0.039+0.026+0.057+0.456+1.379}_{-0.019-0.040-0.117-0.500-0.698}{\times}10^{-4}$
   & $1.415^{+0.039+0.026+0.057+0.455+1.384}_{-0.019-0.040-0.118-0.501-0.699}{\times}10^{-4}$
   & $0.322^{+0.004+0.016+0.006+0.056+0.244}_{-0.004-0.021-0.040-0.108-0.145}{\times}10^{-4}$
   \\ \hline
     $B_{u}^{+}{\eta}^{\prime}$
   & $4.183^{+0.339+0.494+0.470+4.165+6.577}_{-0.040-0.420-0.359-2.034-2.414}{\times}10^{-6}$
   & $4.184^{+0.340+0.494+0.470+4.156+6.576}_{-0.039-0.421-0.362-2.017-2.411}{\times}10^{-6}$
   & $7.225^{+0.082+0.356+0.135+1.261+5.465}_{-0.100-0.476-0.895-2.415-3.256}{\times}10^{-6}$
   \\ \hline
     $B_{u}^{+}K^{0}$
   & $6.334^{+0.033+0.151+0.396+2.218+6.196}_{-0.055-0.289-0.362-1.993-2.853}{\times}10^{-6}$
   & & \\ \hline
     $B_{u}^{+}K^{{\ast}0}$
   & $5.622^{+0.061+0.512+0.675+3.759+7.428}_{-0.103-0.108-0.196-1.491-2.563}{\times}10^{-7}$
   & & \\ \hline \hline
  \end{tabular}
  \end{sidewaystable}

 \begin{figure}[ht]
 \includegraphics[width=0.65\textwidth,bb=190 555 400 590]{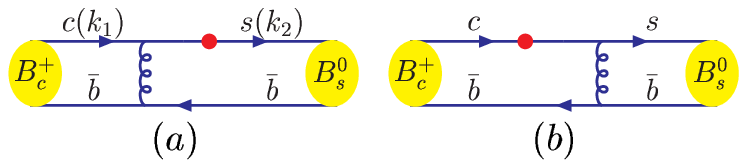}
 \caption{The lowest order diagrams contributing to the
    $B_{c}$ ${\to}$ $B_{s}$ transition form factors, where
    the dot denotes an appropriate Dirac matrix.}
 \label{fig01}
 \end{figure}

 \begin{figure}[ht]
 \includegraphics[width=0.95\textwidth,bb=85 470 520 620]{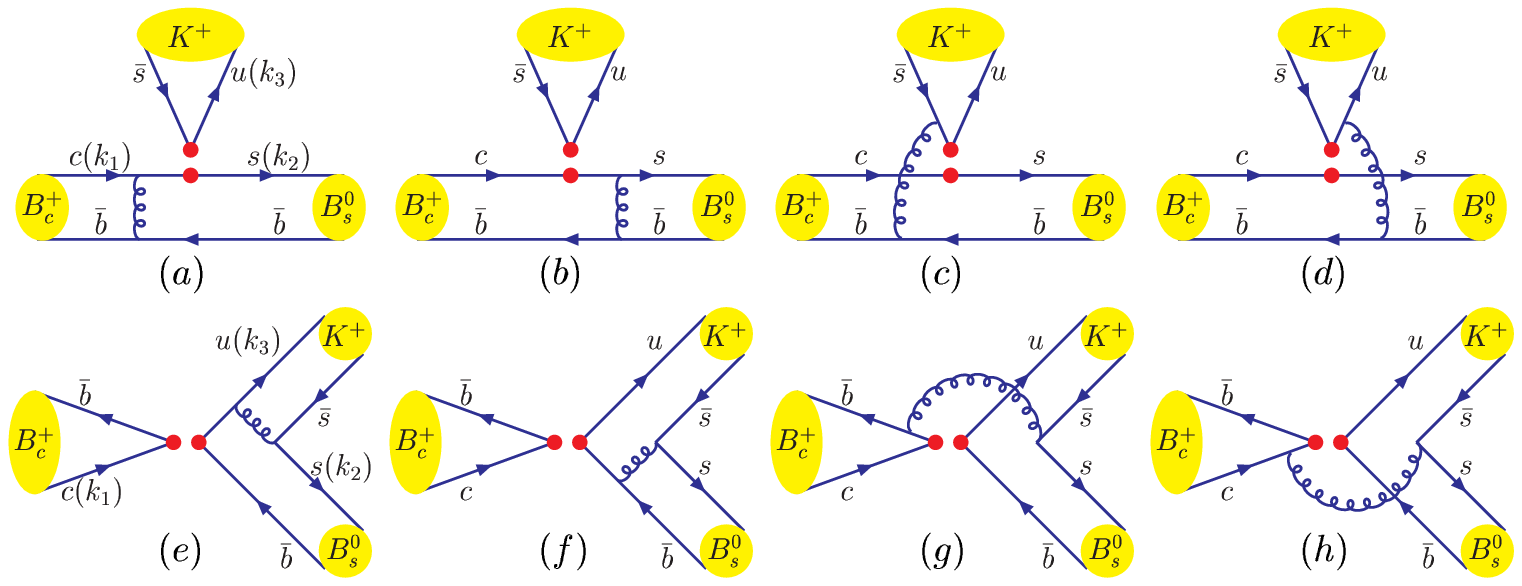}
 \caption{Diagrams contributing to the $B_{c}$ ${\to}$ $B_{s}K$
          decay, where (a) and (b) are called as the factorizable
          emission diagrams;
          (c) and (d) the nonfactorizable emission diagrams;
          (e) and (f) the factorizable annihilation diagrams;
          (g) and (h) the nonfactorizable annihilation diagrams.}
 \label{fig02}
 \end{figure}

 \begin{figure}[ht]
 \includegraphics[width=0.6\textwidth,bb=190 640 380 770]{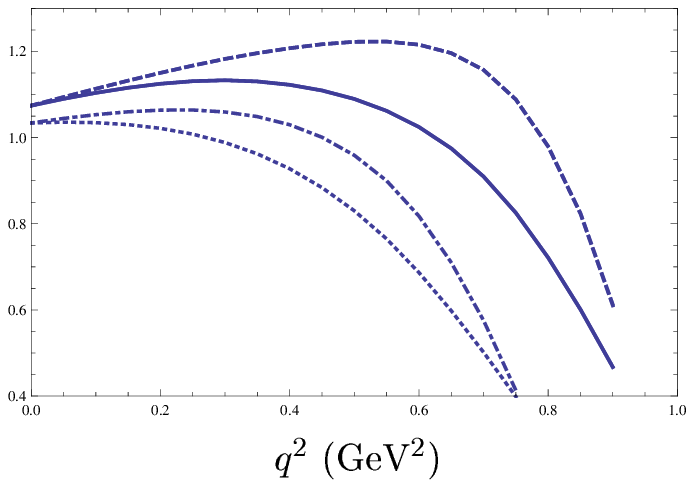}
 \caption{The $q^{2}$ dependence of the form factor, where the
  solid, dashed, dotted, and dotdashed lines denote the
  $F_{0}^{B_{c}{\to}B_{u,d}}(q^{2})$,
  $F_{1}^{B_{c}{\to}B_{u,d}}(q^{2})$,
  $F_{0}^{B_{c}{\to}B_{s}}(q^{2})$ and
  $F_{1}^{B_{c}{\to}B_{s}}(q^{2})$,
  respectively.}
 \label{fig03}
 \end{figure}

 \begin{figure}[ht]
 \includegraphics[width=0.95\textwidth,bb=40 640 550 790]{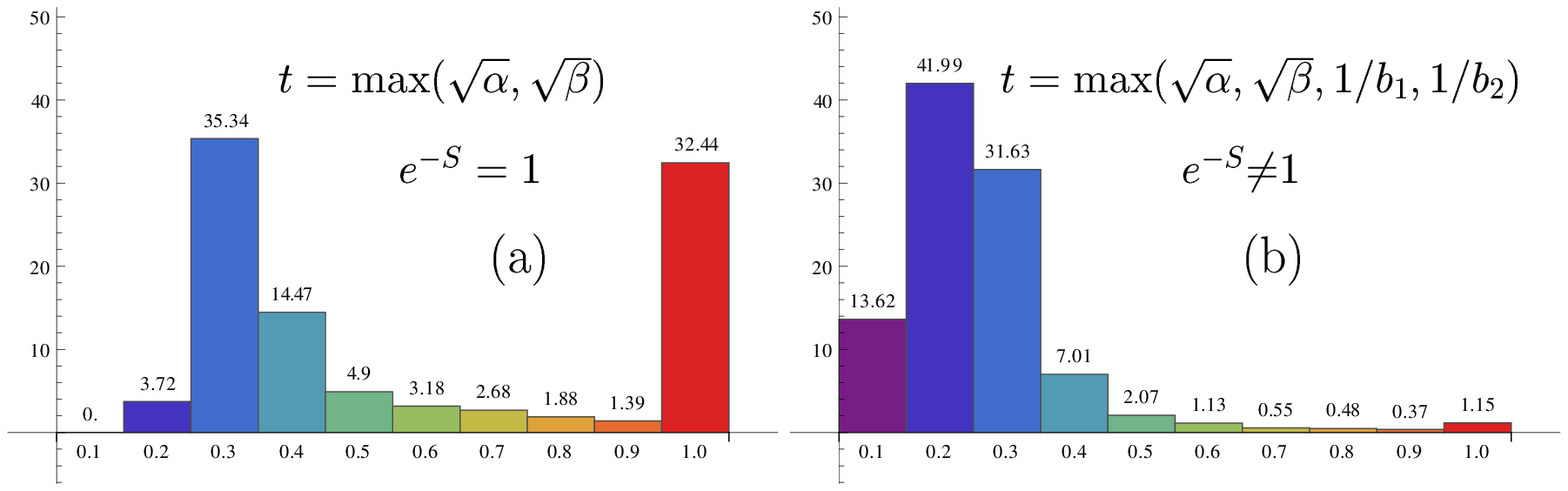}
 \caption{The contributions to the form factor
 $F_{0}^{B_{c}{\to}B_{s}}(0)$ from different
 ranges of ${\alpha}_{s}/{\pi}$,
 where the numbers over histogram denote the
 percentage of the corresponding contributions.}
 \label{fig04}
 \end{figure}

 \begin{figure}[ht]
 \includegraphics[width=0.6\textwidth,bb=85 490 540 780]{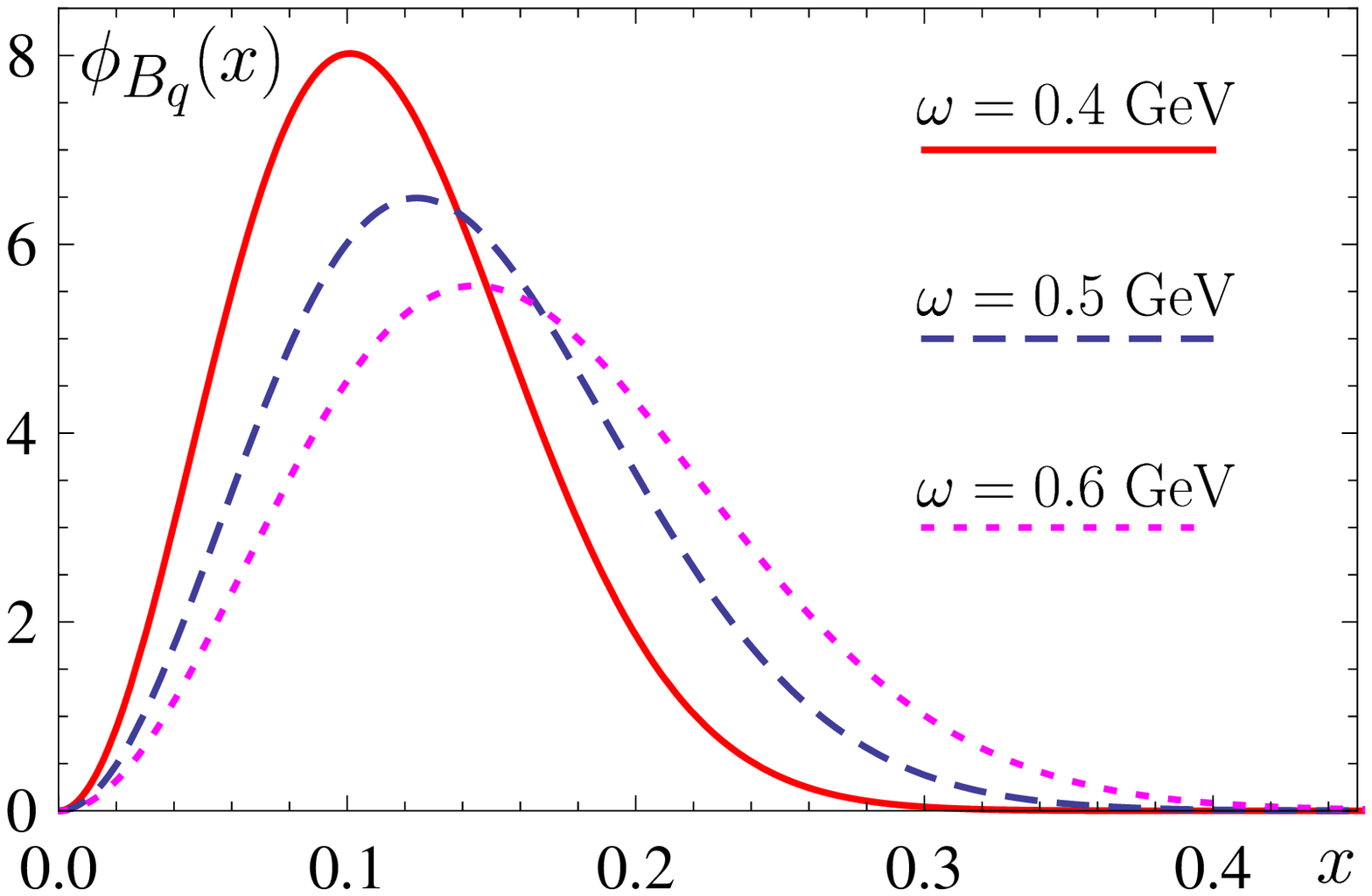}
 \caption{$B_{q}$ meson distribution amplitudes.}
 \label{fig06}
 \end{figure}
 \begin{figure}[ht]
 \includegraphics[width=0.6\textwidth,bb=85 495 540 780]{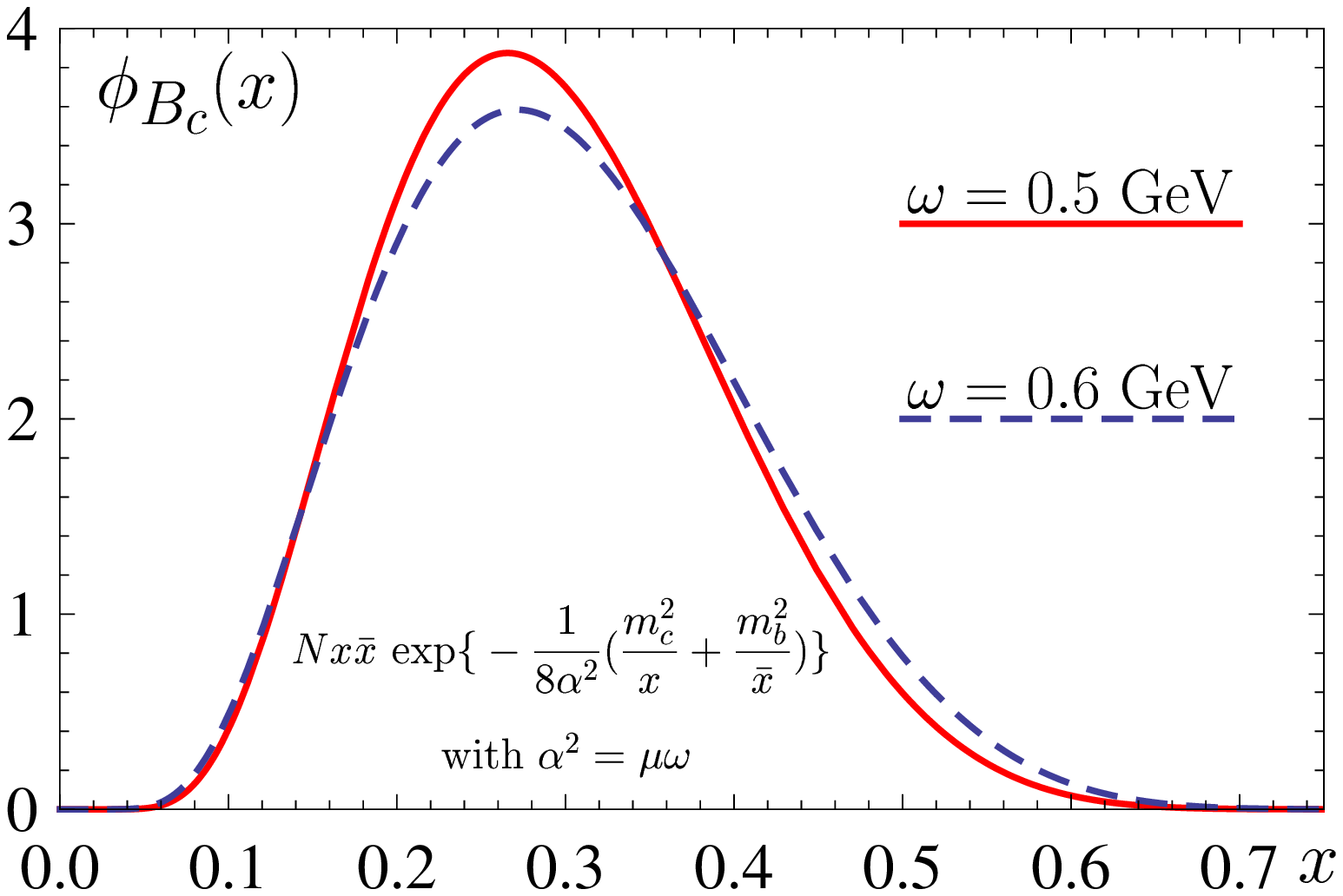}
 \caption{$B_{c}$ meson distribution amplitudes.}
 \label{fig07}
 \end{figure}


\begin{thebibliography}{99}
 \bibitem{pdg.2012}
     J. Beringer et al. (Particle Data Group),
     Phys. Rev. D86, 010001 (2012).
 \bibitem{prd.58.112004}
     F. Abe et al. (CDF Collaboration),
     Phys. Rev. D58, 112004 (1998);
     Phys. Rev. Lett. 81, 2432 (1998).
 \bibitem{prl.100.182002}
     T. Aaltonen et al. (CDF Collaboration),
     Phys. Rev. Lett. 100, 182002 (2008).
 \bibitem{prl.109.232001}
     R. Aaij et al. (LHCb Collaboration),
     Phys. Rev. Lett. 109, 232001 (2012).
 \bibitem{arxiv:1401.6932}
     R. Aaij et al. (LHCb Collaboration),
     arXiv:1401.6932.
 \bibitem{zpc.51.549}
     M. Lusignoli, M. Masetti,
     Z Phys C51, 549 (1991).
 \bibitem{prd.49.3399}
     C. Chang, Y. Chen,
     Phys. Rev. D49, 3399 (1994).
 \bibitem{prd.53.4991}
     M. Beneke, G. Buchalla,
     Phys. Rev. D53, 4991 (1996);
     C. Chang et al.,
     Phys. Rev. D64, 014003 (2001);
     Commun. Theor. Phys. 35, 57 (2001).
 \bibitem{1308.4544}
     R. Aaij et al. (LHCb collaboration),
     Phys. Rev. Lett. 111, 181801 (2013).
 \bibitem{0412158}
     N. Brambilla et al. (Quarkonium Working Group),
     CERN-2005-005, arXiv:hep-ph/0412158.
 \bibitem{prl.101.012001}
     V. Abazov et al. (D0 collaboration),
     Phys. Rev. Lett. 101, 012001 (2008).
 \bibitem{1303.1737}
     R. Aaij et al. (LHCb collaboration),
     Phys. Rev. D87, 071103 (2013).
 \bibitem{prd.87.112012}
     R. Aaij et al. (LHCb Collaboration),
     Phys. Rev. D87, 112012 (2013).
 \bibitem{1309.0587}
     R. Aaij et al. (LHCb collaboration),
     arXiv:1309.0587.
 \bibitem{1204.0079}
     R. Aaij et al. (LHCb collaboration),
     Phys. Rev. Lett. 108, 251802 (2012).
 \bibitem{prl.97.012002}
     A. Abulencia et al. (CDF Collaboration),
     Phys. Rev. Lett. 97, 012002 (2006).
 \bibitem{prd.81.074012}
     S. Descotes-Genon et al.,
     Phys. Rev. D80, 114031 (2009);
     X. Liu, Z. Xiao, C. L\"{u},
     Phys. Rev. D81, 014022 (2010);
     Y. Yang, J. Sun, N. Wang,
     Phys. Rev. D81, 074012 (2010).
 \bibitem{pan.67.1559}
     A. Likhoded, A. V. Luchinsky,
     Phys. Rev. D82, 014012 (2010);
     I. Gouz et al.,
     Phys. Atom. Nucl. 67, 1559 (2004);
     V. Kiselev,
     arXiv:hep-ph/0211021.
 \bibitem{prd.86.094028}
     S. Naimuddin et al.,
     Phys. Rev. D86, 094028 (2012).
 \bibitem{prd.80.114003}
     H. Choi, C. Ji,
     Phys. Rev. D80, 114003 (2009).
 \bibitem{prd.74.074008}
     E. Hern\'{a}ndez, J. Nieves, J. Verde-Velasco,
     Phys. Rev. D74, 074008 (2006).
 \bibitem{prd.73.054024}
     M. Ivanov, J. K\"{o}rner, P. Santorelli,
     Phys. Rev. D73, 054024 (2006);
     Phys. Rev. D63, 074010 (2001).
 \bibitem{epjc.32.29}
     D. Ebert, R. Faustov, V. Galkin,
     Eur. Phys. J. C32, 29 (2003).
 \bibitem{prd.62.014019}
     A. El-Hady, J. Munoz, J. Vary,
     Phys. Rev. D62, 014019 (2000).
 \bibitem{prd.39.1342}
     D. Du, Z. Wang,
     Phys. Rev. D39, 1342 (1989).
 \bibitem{prd.61.034012}
     P. Colangelo, F. Fazio,
     Phys. Rev. D61, 034012 (2000).
 \bibitem{prd.77.114004}
     J. Sun et al.,
     Phys. Rev. D77, 114004 (2008).
 \bibitem{pqcd}
     C. Chang,  H. Li,
     Phys. Rev. D55, 5577 (1997);
     T. Yeh, H. Li,
     Phys. Rev. D56, 1615 (1997);
     Y. Keum, H. Li, A. Sanda,
     Phys. Lett. B504, 6 (2001);
     Y. Keum, H. Li,
     Phys. Rev. D63, 074006 (2001);
     C. L\"{u}, K. Ukai, M. Yang,
     Phys. Rev. D63, 074009 (2001);
     C. L\"{u}, M. Yang,
     Eur. Phys. J. C23, 275 (2002).
 \bibitem{9512380}
      For a review, see
      G. Buchalla, A. Buras, M. Lautenbacher,
      Rev. Mod. Phys. {\bf 68}, 1125, (1996); 
      or A. J. Buras, hep-ph/9806471.
 \bibitem{prd.22.2157}
      G. Lepage, S. Brodsky,
      Phys. Rev. D22, 2157 (1980).
 \bibitem{qcdf}
      M. Beneke et al.,
      Phys. Rev. Lett. 83, 1914 (1999);
      Nucl. Phys. B591, 313 (2000);
      Nucl. Phys. B606, 245 (2001);
      D. Du, D. Yang, G. Zhu,
      Phys. Lett. B488, 46 (2000);
      Phys. Lett. B509, 263 (2001);
      Phys. Rev. D64, 014036 (2001).
 \bibitem{scet}
      C. Bauer, S. Fleming, M. Luke,
      Phys. Rev. D63, 014006 (2001);
      C. Bauer et al.,
      Phys. Rev. D63, 114020 (2001);
      C. Bauer, I. Stewart,
      Phys. Lett. B516, 134 (2001);
      C. Bauer, D. Pirjol, I. Stewart,
      Phys. Rev. D65, 054022 (2002);
      C. Bauer, et al.,
      Phys. Rev. D66, 014017 (2002);
      M. Beneke et al.,
      Nucl. Phys. B643, 431 (2002);
      M. Beneke, T. Feldmann,
      Phys. Lett. B553, 267 (2003);
      Nucl. Phys. B685, 249 (2004).
 \bibitem{prd.65.014007}
      T. Kurimoto, H. Li, A. Sanda,
      Phys. Rev. D65, 014007 (2001).
 \bibitem{bsw}
      M. Wirbel, B. Stech, and M. Bauer,
      Z. Phys. C29, 637 (1985).
 \bibitem{prd.58.114006}
     Th. Feldmann, P. Kroll, B. Stech,
     Phys. Rev. D58, 114006 (1998).
 \bibitem{prd.76.074018}
     A. Ali et al.,
     Phys. Rev. D76, 074018 (2007).
 \bibitem{JHEP.05.006.2007}
     R. Escribano, J. Nadal,
     JHEP 05, 006 (2007);
     F. Ambrosino et al. (KLOE collaboration),
     JHEP 07, 105 (2009).
 \bibitem{prd.87.097501}
     W. Wang et al.,
     Phys. Rev. D87, 097501 (2013);
     Y. Charng et al.,
     Phys. Rev. D74, 074024 (2006).
 \bibitem{epjc.28.515}
     C. L\"{u}, M. Yang,
     Eur. Phys. J. C28, 515 (2003).
 \bibitem{1310.8555}
     S. Aoki et al. (FLAG Working Group),
     arXiv:1310.8555 [hep-lat],
     http://itpwiki.unibe.ch/flag
 \bibitem{JHEP.03.069.2007}
     P. Ball, G. Jones,
     JHEP 03, 069 (2007).
 \bibitem{plb.651.171}
     T. Chiu et al. (TWQCD collaboration),
     Phys. Lett. B651, 171 (2007).
 \bibitem{JHEP.05.004.2006}
     P. Ball, V. Braun, A. Lenz,
     JHEP 05, 0004 (2006).
 \bibitem{Phys.Rev.D63.054008}
     Y. Keum, H. Li, A. Sanda,
     Phys. Rev. D63, 054008 (2001).
 \bibitem{omega}
         D. Ebert, R. Faustov, V. Galkin,
         Eur. Phys. J. C71, 1825 (2011);
         B. Patel, P. Vinodkumar,
         J. Phys. G36, 035003 (2009)
         and references therein.
 \bibitem{Phys.Rev.D22.2157}
         G. Lepage, S. Brodsky,
         Phys. Rev. D22, 2157 (1980);
         T. Huang, B. Ma, Q. Shen,
         Phys. Rev. D49, 1490 (1994).
 \bibitem{JHEP.01.010.1999}
         P. Ball, JHEP 01, 010 (1999).
 \bibitem{Nucl.Phys.B529.323}
         P. Ball et al.,
         Nucl. Phys. B529, 323 (1998).
 \bibitem{Phys.Rev.D52.3958}
         H. Li, Phys. Rev. D52, 3958 (1995).
 \end{thebibliography}
  \end{document}